\documentclass[aps,prl,notitlepage,twocolumn,nofootinbib,superscriptaddress,longbibliography]{revtex4-1}

\usepackage{amssymb,amsmath,amsfonts,bm,color,graphicx,multirow}
\usepackage[linktocpage]{hyperref}
\hypersetup{colorlinks=true,citecolor=blue,linkcolor=blue,filecolor=blue,urlcolor=blue,breaklinks=true}
\newcommand{\nc}{\newcommand}
\nc{\ket}[1]{|#1\rangle}
\nc{\bra}[1]{\langle#1|}
\nc{\ketbra}[2]{|#1\rangle\!\langle#2|}
\nc{\braket}[2]{\langle#1|#2\rangle}
\nc{\braoprket}[3]{\langle#1|#2|#3\rangle}
\nc{\avg}[1]{\langle#1\rangle}
\nc{\ketbrasame}[1]{|#1\rangle\!\langle#1|}
\nc{\tr}{\operatorname{tr}}

\usepackage{tikz}
\definecolor{darkblue}{RGB}{0,76,156}
\definecolor{darkred}{RGB}{195,0,0}
\nc{\ZY}[1]{{\color{darkblue}{[ZY: #1]}}}
\nc{\RY}[1]{{\color{magenta}{[RY: #1]}}}
\nc{\HK}[1]{{\color{violet}{[HK: #1]}}}

\begin{document}
\title{Pair density wave characterized by a hidden string order parameter}

\author{Hao-Kai Zhang}
\affiliation{Institute for Advanced Study, Tsinghua University, Beijing 100084, China}
\author{Rong-Yang Sun}
\affiliation{Computational Materials Science Research Team, RIKEN Center for Computational Science (R-CCS), Kobe, Hyogo 650-0047, Japan}
\affiliation{Quantum Computational Science Research Team, RIKEN Center for Quantum Computing (RQC), Wako, Saitama 351-0198, Japan}
\author{Zheng-Yu Weng}
\affiliation{Institute for Advanced Study, Tsinghua University, Beijing 100084, China}

\date{\today}

\begin{abstract}
A composite pairing structure of superconducting state is revealed by density matrix renormalization group study in a two-leg $t$-$J$ model. The pairing order parameter is composed of a pairing amplitude and a phase factor, in which the latter explicitly depends on the spin background with an analytic form identified in the anisotropic limit as the interchain hopping integral $t_{\perp}\rightarrow 0$. Such a string-like phase factor is responsible for a pair density wave (PDW) induced by spin polarization with a wavevector $Q_{\mathrm {PDW}}=2\pi m$ ($m$ the magnetization). By contrast, the pairing amplitude remains smooth, unchanged by the PDW. In particular, a local spin polarization can give rise to a sign change of the order parameter across the local defect. Unlike in an Fulde-Ferrell-Larkin-Ovchinnikov state, the nonlocal phase factor here plays a role as the new order parameter characterizing the PDW, whose origin can be traced back to the essential sign structure of the doped Mott insulator.
\end{abstract}

\maketitle

\textit{Introduction.---} Pair density wave (PDW) states are superconducting (SC) states in which Cooper pairs have a finite center-of-mass momentum so that the SC order parameter oscillates spatially with vanishing average~\cite{*[{}] [{ and reference therein.}] Agterberg2020}. Such states have been studied to understand cuprate superconductors especially for the dynamical inter-layer decoupling phenomena~\cite{Himeda2002,Li2007,Berg2007,Lozano2022}. Signatures of PDW states have also been reported via local Cooper pair tunneling and scanning tunneling microscopy~\cite{Hamidian2016,Ruan2018,Edkins2019}. In weak correlated systems, the Fulde-Ferrell-Larkin-Ovchinnikov (FFLO) state~\cite{Fulde1964,Larkin1965} is the first example of PDW states based on the BCS theory where the Fermi surfaces of different spins are split by the Zeeman field. By contrast, in strongly correlated systems such as doped Mott insulators where the standard BCS picture may not generally hold, the mechanism for PDW is still under debate in either the presence or absence of an external magnetic field \cite{Berg2009,Berg2010,Loder2010,Loder2011,Lee2014,Wardh2017,Wardh2018,Setty2021,Setty2022,Jiang2022,Wu2022}.

To explore the mechanism of superconductivity in a doped Mott insulator, a two-leg $t$-$J$ ladder may serve as an interesting toy model~\cite{Poilblanc1995,Roux2006,Jiang2020,Sun2020,Shinjo2021}, in which a quasi-one-dimensional SC state or the Luther-Emery (LE) liquid~\cite{Luther1974} has been previously established, with a strong pairing of doped holes in a short-range antiferromagnetic (AFM) spin background. In particular, the LE phase remains robust and can be smoothly extrapolated to the limit of the inter-chain hopping integral $t_{\perp}=0$ \cite{Jiang2020}. In the latter, an analytic composite pairing structure can be identified, where the pairing amplitude and phase is explicitly separated to describe the pairing of the ``twisted'' doped holes and a nonlocal spin-dependent phase shift, respectively \cite{Zhu2018,Chen2018a,Jiang2020}. It suggests that SC phase coherence may be critically examined by fine-tuning the spin background without destroying the pairing at a given doping concentration.

\begin{figure}[b]
	\centering
	\includegraphics[width=0.88\linewidth]{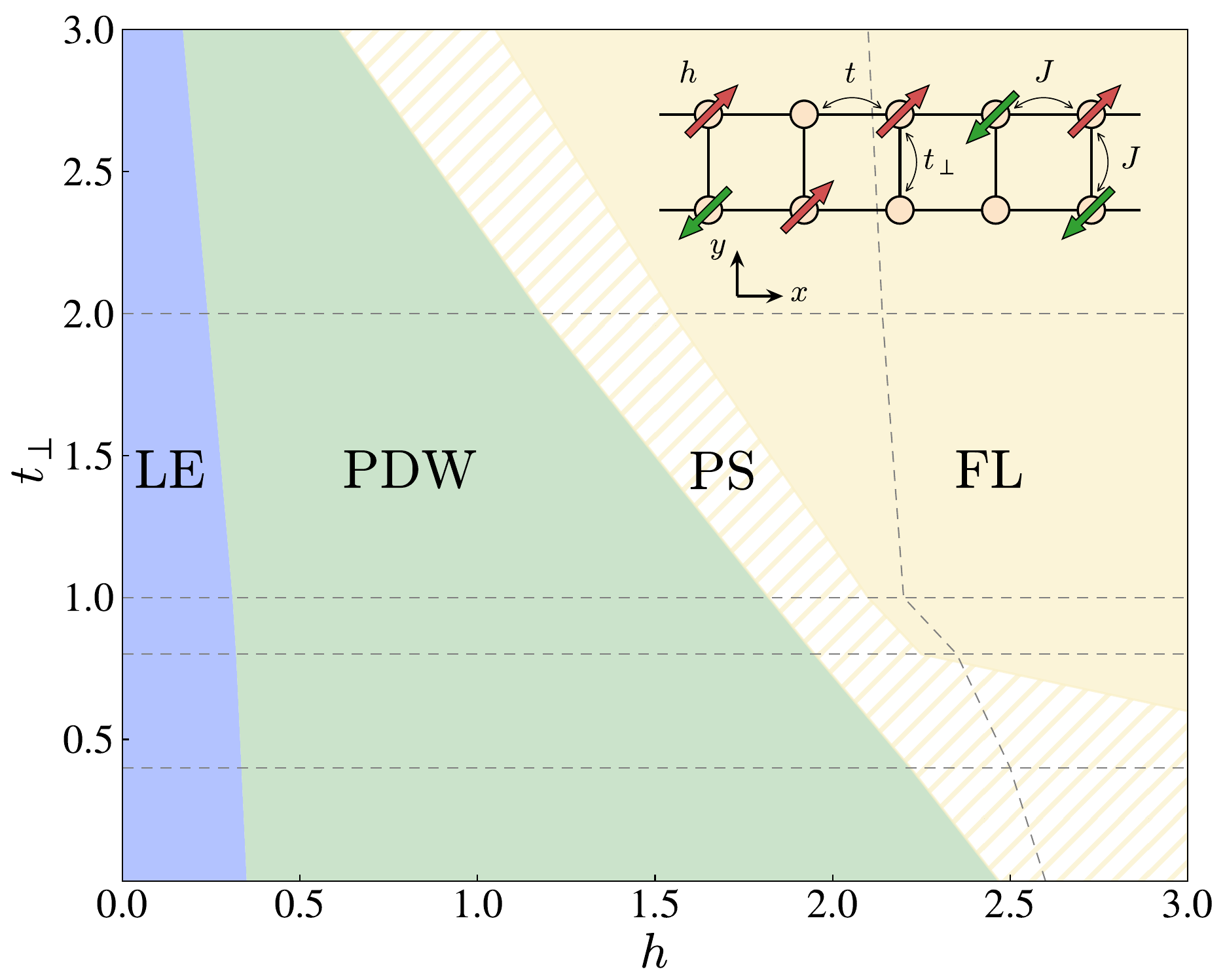}
	\caption{
		Ground state phase diagram of two-leg $t$-$J$ ladder (see the inset) at $t/J=3$ and $\delta=1/8$ with respect to the Zeeman field $h$ and the interchain hopping $t_\perp$ in units of $J$. The three colored regions are identified as the LE, PDW and FL phases, with the shaded area as the phase separation (PS) region. The horizontal and vertical dashed lines mark the swept parameter points and the magnetization saturation, respectively~\cite{SM}.
	}
	\label{fig:phase_diagram}
\end{figure}

In this paper, we employ the density matrix renormalization group (DMRG) method~\cite{White1992} to study the ground state of the doped $t$-$J$ two-leg ladder by polarizing the background spins. It is shown that the pairing remains robust beyond the LE phase to persist into a PDW state once a uniform spin magnetization ($m$) sets in, resulting in an algebraic-decaying SC correlator oscillating in sign spatially. In particular, the pairing order parameter can change sign even across locally polarized spins, indicating a non-FFLO type of mechanism. The hole pairing eventually vanishes and a Fermi liquid (FL) occurs as schematically summarized in Fig.~\ref{fig:phase_diagram}. In the limit of $t_{\perp}\rightarrow 0$, an analytic form enables an explicit separation of a pairing amplitude from a string-like phase factor, which plays a role of an order parameter to solely determine the PDW wavevector $Q_{\mathrm {PDW}}=2\pi m$. Finally it is argued generally that the PDW is a direct manifestation of the phase-string sign structure in the model, and the PDW disappears in the whole phase diagram once the phase-string is turned off in the DMRG simulation.

\begin{figure}
	\includegraphics[width=\linewidth]{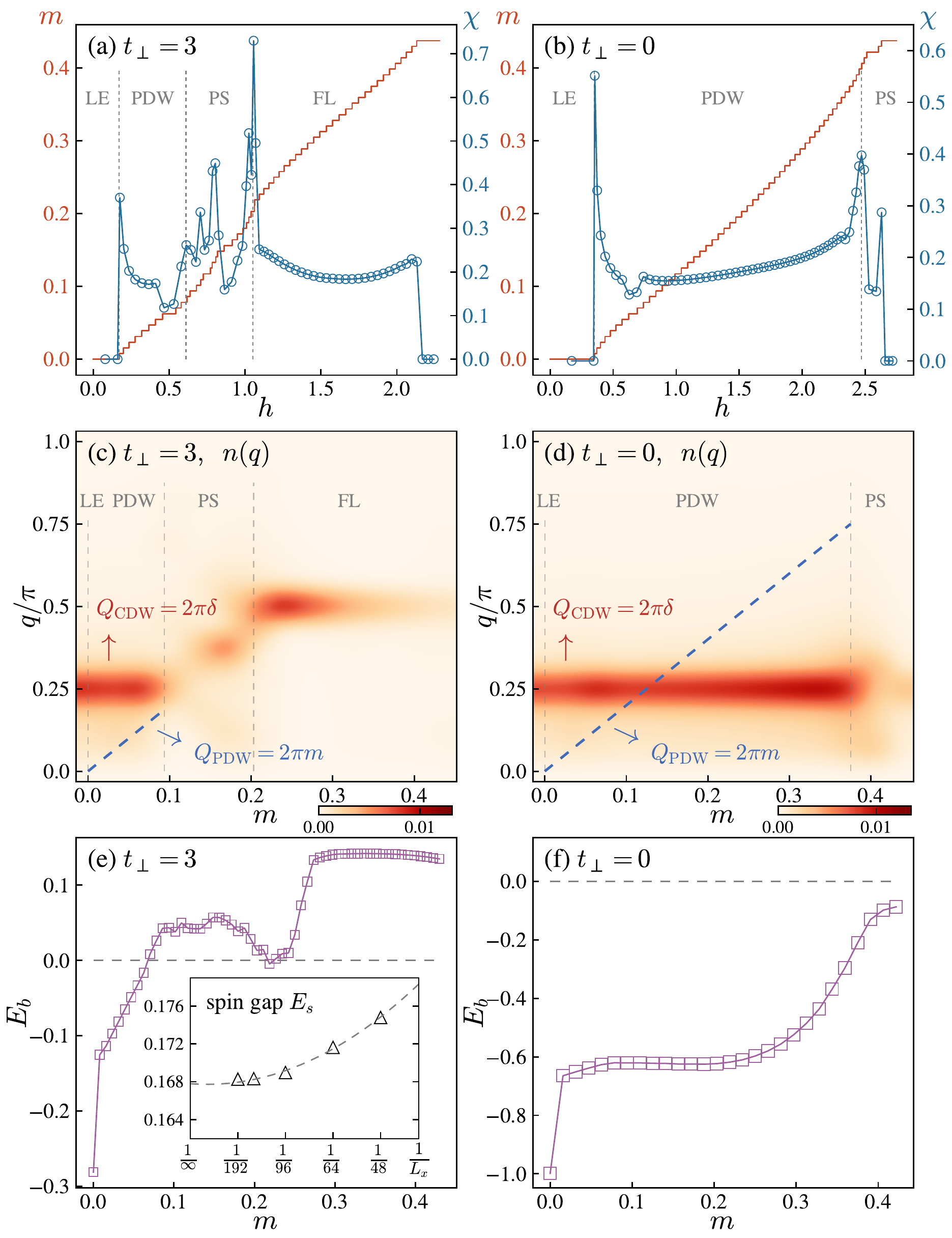}
	\caption{(a) and (b): Magnetization $m$ and static susceptibility $\chi$ versus the Zeeman field $h$ for the isotropic limit $t_{\perp}=3$ and anisotropic limit $t_{\perp}=0$ of a two-leg $t$-$J$ ladder with $L_x=64$ at $\delta=1/8$.  (c) and (d): Fourier spectrum $n(q)$ of the charge density versus $m$. The CDW wavevector $Q_{\rm CDW}$ and PDW wave vector $Q_{\rm PDW}$ are independent from each other in the LE and PDW phases. (e) and (f): Binding energy $E_b$ of two doped holes as a function of $m$ (with $L_x=64,32$), where negative values of $E_b$ indicate the existence of pairing. The inset in (e) depicts the finite-size scaling of the spin gap $E_s$ for the isotropic case.}
	\label{fig:nq_Eb_mh_tp03}
\end{figure}

\textit{Model and method.---} The basic physics of doped Mott insulators in the strong coupling limit is described by the standard $t$-$J$ model with Hamiltonian $H_{t\text{-}J} = \mathcal{P}_s \left(H_t + H_J\right) \mathcal{P}_s$, where 
\begin{equation}\label{Eq:tJ_hamiltonian}
\begin{aligned}
    H_t &= - \sum_{\langle ij\rangle \sigma} t_{ij} \left( c^{\dagger}_{i\sigma} c_{j\sigma} + h.c.\right), \\
    H_J &= \sum_{\langle ij\rangle} J_{ij } \left (\mathbf{S}_i\cdot \mathbf{S}_j - \frac{1}{4} n_i n_j \right).
\end{aligned}
\end{equation}
Here $c_{i\sigma}$ is the electron annihilation operator on lattice site $i=(x,y)$ with spin index $\sigma\in\{\uparrow,\downarrow\}$. $\mathbf{S}_i$ and $n_i$ are the spin and electron number operators, respectively. The Hilbert space is constrained by the no-double-occupancy condition $n_i=\sum_{\sigma} n_{i\sigma} = \sum_{\sigma} c^\dagger_{i\sigma} c_{i\sigma} \leq 1$ on each site imposed by the projector $\mathcal{P}_s$. $t_{ij}$ is the hopping integral and $J_{ij}$ is the superexchange coupling between the nearest-neighbor (NN) sites $\langle ij \rangle$ on a square lattice of size $N=L_{x}\times L_{y}$. We fix $J_{ij}=J=1$ as the energy unit and $t=3$ along the $\hat{x}$-direction. The interchain hopping integral along the $\hat{y}$-direction is set to be $t_{ij}=t_{\perp}$ as an adjustable parameter. To polarize the background spins, we apply a uniform Zeeman field via $H_h = - h S^z$, where $S^z=\sum_i S^z_i$ denotes the total spin $z$-component. Here we focus on two-leg ladders $L_y=2$ with length $L_{x}$ up to $192$ and hole doping at $\delta = N_{h} / N = 1/8$ where $N_{h}$ denotes the number of doped holes. We perform $50$-$200$ sweeps and keep the bond dimension up to $15000$ with a typical truncation error $\epsilon\lesssim 10^{-8}$. The simulations are based on the GraceQ project~\cite{graceq}.

\textit{Phase diagram.---} Three distinct phases are identified for the two-leg $t$-$J$ ladder at finite doping by tuning the spin polarization $m\equiv \langle S^z \rangle/N$ via the Zeeman field $h$ (see Fig.~\ref{fig:phase_diagram}). At $m=0$, the ground state is a robust LE liquid with a quasi-long-range SC order in the whole region of $0\leq t_{\perp}/t \leq 1$ \cite{Jiang2020}. A phase transition occurs at $m\neq 0$ to a PDW phase, which is characterized by an algebraic-decaying SC correlation that oscillates in sign (see below). Eventually a Fermi liquid (FL) phase sets in, which can be continuously extrapolated to the spin fully polarized limit. A phase separation (PS) region lies between the PDW and FL phases, which are distinguished by the pairing and non-pairing of the doped holes. 

The magnetization $m$ is shown in Fig.~\ref{fig:nq_Eb_mh_tp03}(a) and (b) together with the static susceptibility $\chi=\frac{\partial m}{\partial h}$ at $t_{\perp}/t = 1$ and $ t_{\perp}=0$, respectively. The LE/PDW phase boundary is at the onset of $m$ (cf. the sharp peak of $\chi$ on the weak field side) where the Zeeman energy equals to the spin gap: i.e., $h \langle S^z\rangle =E_s$, with $E_s$ shown in the inset of Fig.~\ref{fig:nq_Eb_mh_tp03}(e) which is defined by
\begin{equation}
    E_s(N_h) = E(N_h,S^z=1)-E(N_h,S^z=0)~.
\end{equation}
Here $E(N_h,S^z)$ denotes the ground state energy with the fixed number of doped holes $N_h$ and the total spin $S^z$. 

The LE phase has a CDW or charge density modulation at wavevector $Q_{\rm CDW}=2\pi \delta$ \cite{Jiang2020}, which persists over to the whole PDW phase as indicated in Fig.~\ref{fig:nq_Eb_mh_tp03}(c) and (d). Here the charge density distribution along the $\hat{x}$-direction is defined by $n(x)=\sum_{y=1}^{L_y} \langle n_{x,y} \rangle / L_y$ with the Fourier transformation $n(q)=\left|\sum_{x}n(x)e^{i q x}\right|/R$ (here $x$ is summed over the central bulk of length $R=L_x/2$ to reduce the boundary effect). 

In sharp contrast, $Q_{\rm CDW}$ is independent of the PDW wavevector at $Q_{\rm PDW}=2\pi m$ (see below). Both of which disappear simultaneously at the boundary of the phase separation, indicated in Figs.~\ref{fig:nq_Eb_mh_tp03}(c) and (d), where the density profile becomes spatially inhomogeneous like a mixture of the PDW and FL regimes on the both sides. One is referred to Fig.~\textcolor{blue}{S1} of the Supplemental Material~\cite{SM} for more details. 

Furthermore, the hole pairing can be determined by computing the binding energy
\begin{equation}\label{Eq:Eb}
\begin{aligned}
    E_{b}&(N_h,S^z) \equiv E(N_h+2,S^z)+E(N_h,S^z)\\
    &-E(N_h+1,S^z+\frac{1}{2})-E(N_h+1,S^z-\frac{1}{2})~.
\end{aligned}
\end{equation}
As shown in Figs.~\ref{fig:nq_Eb_mh_tp03}(e) and (f), $E_b$ is negative in both the LE and PDW phases, which then vanishes and becomes positive in the phase separation and FL regions.

\textit{PDW and composite pairing structure.---} The PDW phase has been previously conjectured to be an FFLO-like state in the isotropic limit~\cite{Roux2006}. Nevertheless, the independence of $Q_{\rm CDW}$ and $Q_{\rm PDW}$ [cf. Fig.~\ref{fig:nq_Eb_mh_tp03}(c) and (d)] strongly suggest that both orders may not be originated from a naive spin-polarized Fermi surface effect. According to the phase diagram, the LE and PDW phases can continuously persist over the whole region of $0\leq t_{\perp}/t \leq 1$, and in particular, the PDW is most robust against a finite $m$ at $t_{\perp}=0$ in Fig.~\ref{fig:phase_diagram}. In the following, we shall first focus on the PDW state in the anisotropic limit of $t_{\perp}=0$, where some precise and useful analytic structure is available~\cite{Zhu2018,Chen2018a}.  

\begin{figure}
	\includegraphics[width=\linewidth]{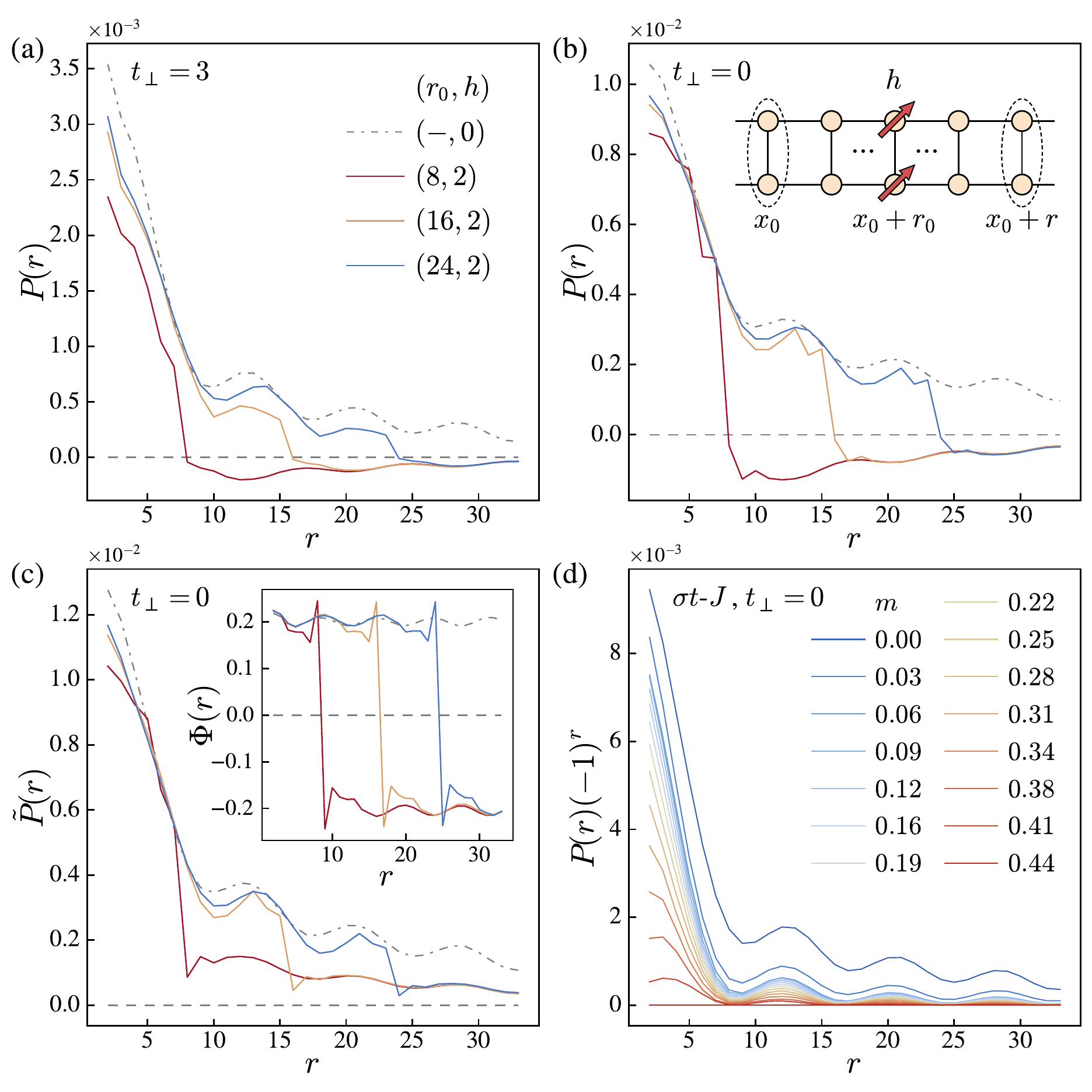}
	\caption{Pair-pair correlator at $t_\perp=3$ (a) and $t_\perp=0$ (b), respectively, shows a sign change across the local spin polarizations at rung $x_0+r_0$, which are induced by a strong local Zeeman field as indicated in the inset of (b). (c) The composite pairing structure: the phase-shift operator (the inset) is responsible for the sign change of the pair-pair correlator, while the pairing amplitude is unchanged in sign across the local ``defect''. (d) By turning off the phase-string in the $t$-$J$ ladder (see text), the pair-pair correlator no longer exhibits any additional ``PDW" order at a finite $m$ ($L_x=64$).
	}
	\label{fig:sc_local_h_sigma}
\end{figure}

Here the pairing order parameter may be explicitly decomposed into the amplitude and phase components~\cite{Zhu2018,Chen2018a} in a form $\avg{\Delta_x}\propto\avg{\tilde{\Delta}_x}\avg{e^{i\phi_{x}}}$. Specifically, the spin-singlet pair operator $\Delta_x\equiv \frac{1}{\sqrt{2}}\sum_{\sigma}\sigma c_{x1,\sigma} c_{x2,-\sigma}$, which is defined at the rung of sites $(x,1)$ and $(x,2)$, may be rewritten as~\cite{SM}
\begin{equation}\label{Eq:cs-separa}
\begin{aligned}
    \Delta_x =& \frac{1}{\sqrt{2}}\Big( \tilde{c}_{x1,\uparrow} \tilde{c}_{x2,\downarrow} e^{i(\Omega_{x1,\downarrow}+\Omega_{x2,\uparrow})} \\
    & - \tilde{c}_{x1,\downarrow} \tilde{c}_{x2,\uparrow}e^{i(\Omega_{x1,\uparrow}+\Omega_{x2,\downarrow})}\Big),
\end{aligned}
\end{equation}
where the ``twisted'' holes are defined by $\tilde{c}_{i,\sigma} = c_{i,\sigma} e^{-i \Omega_{i,-\sigma}}$, with the string operator $\Omega_{xy,\sigma} \equiv \pi \sum_{x'>x} n_{x'y,\sigma}$ in which the summation runs over one of the 1D chains of the two-leg ladder. Define the pairing amplitude $\tilde{\Delta}_x\equiv \frac{1}{\sqrt{2}}\sum_{\sigma}\sigma \tilde{c}_{x1,\sigma} \tilde{c}_{x2,-\sigma}$ and the string-like phase $\phi_x\equiv \Omega_{x1,\downarrow}+\Omega_{x2,\uparrow}$~\cite{SM}. Correspondingly, the pair-pair correlators of $P(r)=\langle\Delta^\dagger_{x_0}\Delta_{x_0+r}\rangle$, as well as the sub-components $\tilde{P}(r)=\langle \tilde{\Delta}^\dagger_{x_0} \tilde{\Delta}_{x_0+r}\rangle$ and $\Phi(r) = \avg{ e^{-i\phi_{x_0}} e^{i\phi_{x_0+r}} }$ can be computed separately. The results including their Fourier transformations are presented in Fig.~\ref{fig:sc_local_h_sigma} (here the rung at $x_0=L_x/4$ is set as a reference bond and the distance $r$ is between two rungs). 

\begin{figure}
	\includegraphics[width=\linewidth]{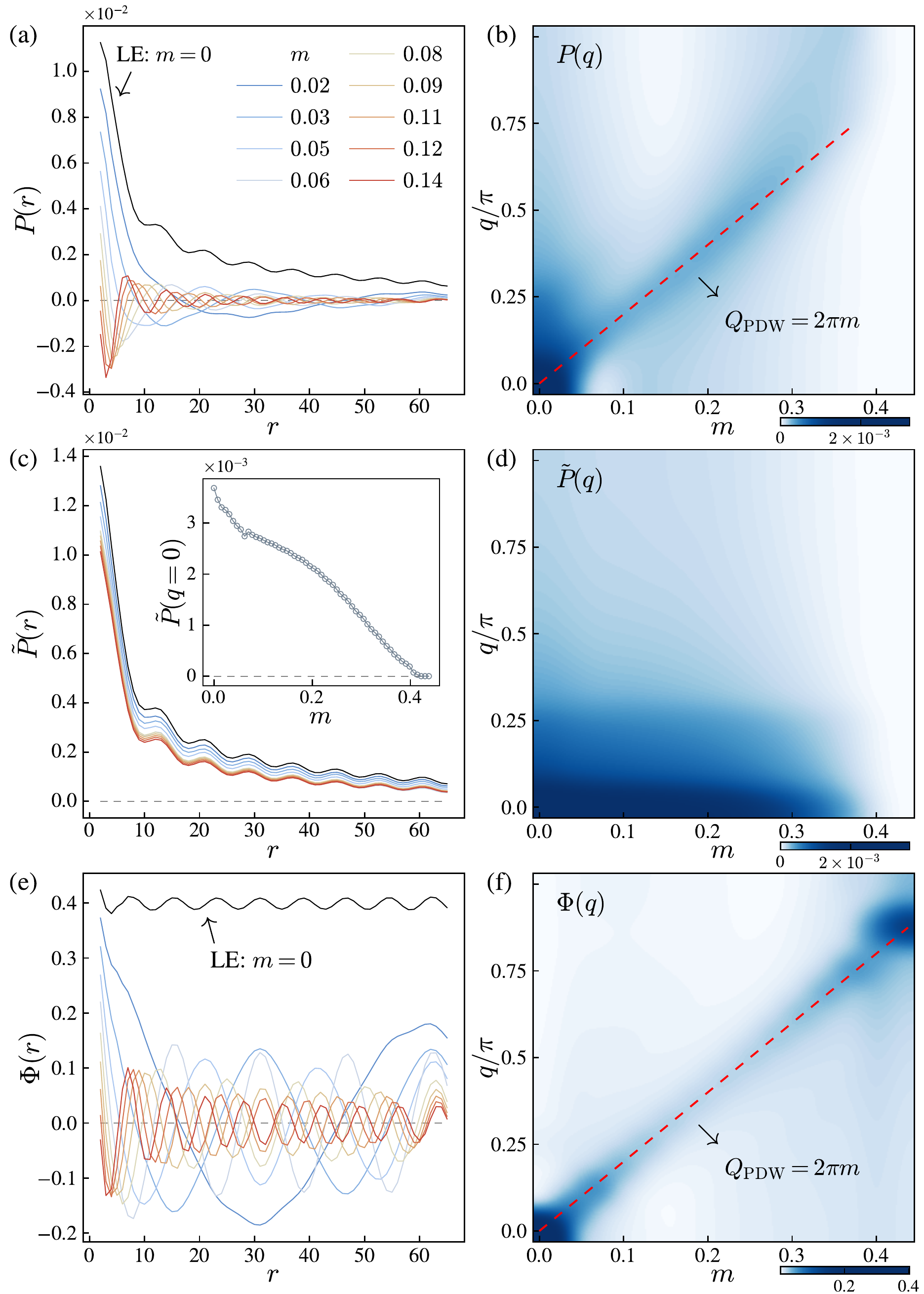}
	\caption{Composite structure of the pairing order parameter at $t_{\perp}=0$ ($\delta=1/8$). (a) and (b): Pair-pair correlator in the real space ($L_x=128$) and momentum space ($L_x=64$) versus $m$, respectively. 
	(c) and (d): The corresponding correlator of the pairing amplitude  (see text). The inset in (c) depicts the overall magnitude of $\tilde{P}(r)$ as a function of $m$.
     (e) and (f): The corresponding correlator of the phase component of the pairing order parameter in the real and momentum space.}
	\label{fig:sc_r_q}
\end{figure}

At $m=0$, the two pair-pair correlators, $P(r)$ and $\tilde{P}(r)$, are essentially the same \cite{Zhu2018} as shown in Fig.~\ref{fig:sc_r_q}(a) and (c), which is due to the fact that the phase fluctuation of the phase $\phi_x$ gets cancelled out by the short-range AFM correlation. But once $m\neq 0$, $P(r)$ exhibits a PDW oscillation at a \emph{new} wavevector $Q_{\rm PDW}=2\pi m$ [Fig.~\ref{fig:sc_r_q}(a) and (b)] with a polynomial-decaying envelope. By contrast, $\tilde{P}(r)$ still behaves the same as in the LE ($m=0$) phase such that the PDW oscillation entirely comes from the phase factor as shown by $\Phi(r)$ in Fig.~\ref{fig:sc_r_q}(e) and (f). The DMRG results shown in Fig.~\ref{fig:sc_r_q} confirm that the SC order parameter can be indeed decomposed into an amplitude $\tilde{\Delta}_x$ and a phase factor $e^{i\phi_{x}}$, respectively, which behave independently in the $t_{\perp}= 0$ limit. Namely, the amplitude component will characterize the preformed hole pairs in both the LE and PDW phases, with $\tilde{P}(r)$ only smoothly reducing with the increase of $m$. On the other hand, the string-like phase factor changes qualitatively from a constant at $m=0$ (with a weak CDW ingredient) to a predominant PDW oscillation at $m\neq 0$ in Fig.~\ref{fig:sc_r_q}(c). In other words, such a string-like phase factor can serve as the \emph{sole} PDW order parameter. As a matter of fact, it is clearly shown in Figs.~\textcolor{blue}{S3} and \textcolor{blue}{S4} of the Supplemental Material~\cite{SM}, that the product of $\tilde{P}(r)$ and $\Phi(r)$ can very precisely reproduce the curve of $P(r)$ except for an overall numerical factor, which means that the composite pairing structure $\avg{\Delta_x}\propto\avg{\tilde{\Delta}_x}\avg{e^{i\phi_{x}}}$ or the separation of the pairing amplitude (charge) and the phase (spin) is indeed
highly accurate in the sense of a generalized mean-field description. A generalization to the two-dimensional isotropic case has been made recently for two holes~\cite{Zhao2022b}.

It is important to note that beyond a uniform Zeeman field, a local spin polarization in the background may also lead to a nonlocal phase change in the SC order parameter via $e^{i\phi_{x}}$. To test this effect, one may apply a strong local Zeeman field to polarize the spins in a given rung at $x_0+r_0$ [cf. the inset of Fig.~\ref{fig:sc_local_h_sigma}(b)] which indeed leads to the sign changes of the SC correlator between rung $x_0$ and $r$ across the defect at $x_0+r_0$ for $t_{\perp}=3$ (isotropic case) and $t_{\perp}=0$, respectively, as illustrated in Fig.~\ref{fig:sc_local_h_sigma}(a) and (b) at different $r_0$'s. In particular, the corresponding pairing amplitude correlator $\tilde{P}(r)$ remains positive-definite, while $\Phi(r)$ changes sign across the defect, which are illustrated in Fig.~\ref{fig:sc_local_h_sigma}(c) at $t_{\perp}=0$.

\textit{Physical origin of PDW.---} We have so far identified a novel entanglement between the Cooper pair and the background spins that can lead to a PDW order via spin polarization. In the following, we give a general physical argument on the origin of such a PDW based on a fundamental property of the doped Mott insulator. We note that the partition function of the $t$-$J$ model can be in general expressed as $Z_{t\text{-}J}=\sum_{c} \tau_c W[c]$ with $W[c]\geq 0$ and $\tau_c = (-1)^{N^{h}_{\downarrow}[c]}(-1)^{N^{h}_{\rm ex}[c]}$ where $c$ represents the closed loops of all the spins and holes~\cite{Wu2008}. Here $(-1)^{N^{h}_{\rm ex}[c]}$ represents a conventional fermionic sign structure for the doped holes like in a doped semiconductor, but $(-1)^{N^{h}_{\downarrow}[c]}$ is unique for a doped Mott insulator, which depends on the parity of the number of \emph{mutual exchanges} between holes and $\downarrow$-spins, known as the phase-string. Noting that $N^{h}_{\downarrow}= \left(N^{h}_{\uparrow}+N^{h}_{\downarrow}\right)/2+\left(N^{h}_{\uparrow}-N^{h}_{\downarrow}\right)/2$. Then it is easy to see that the second term can contribute to a nontrivial spin-polarization-dependent phase factor $\exp \left[\pm i\pi \sum_{i\in c}S^z_i\right] $ associated with each hole path. Accordingly for a pair of holes traversing along the $\hat{x}$-direction of the two-leg ladder, by a distance $L_{\mathrm {AB}}$, the pair-pair correlators will generally acquire an additional phase factor $\exp\left[\pm i 2\pi \langle S^z_{\mathrm {AB}}\rangle \right] $, which gives rise to the PDW wavevector $Q_{\mathrm {PDW}}$ by noting that $\langle S^z_{\mathrm {AB}}\rangle=m L_{\mathrm {AB}}$ for a uniform $m$, or the sign change in the SC order parameter across a sufficiently large local spin polarization [cf. Figs.~\ref{fig:sc_local_h_sigma}(a)-(c)]. 
Finally, the phase-string factor in $\tau_c$ can be precisely switched off, without changing $W[c]$ in $Z_{t\text{-}J}$, by inserting a spin-dependent sign $\sigma$ in the hopping term $H_{\sigma t} = - \sum_{\langle ij\rangle \sigma} \sigma t_{ij}c^{\dagger}_{i\sigma} c_{j\sigma} + h.c.$,
which results in the so-called $\sigma\cdot t$-$J$ model~\cite{Zhu2012}. Then carrying out the similar DMRG calculation, one finds that the PDW is no longer present at $m\neq 0$ as shown in Fig.~\ref{fig:sc_local_h_sigma}(d).

In conclusion, a Cooper pair moving on a spin-polarized background can acquire a new oscillating phase, which illustrates a general mutual entanglement between the doped charge and spin degrees of freedom in the $t$-$J$ model. In contrast to a conventional FFLO state in the BCS theory, the doped holes here are drastically ``twisted'' by the underlying phase-string effect, as explicitly shown in the $t_{\perp}=0$ case, which can result in strongly preformed pairing of holes that is not sensitive to the Zeeman field. However, the phase coherence of the hole pair is further tied to the singlet or resonating-valence-bond pairing of spins. Once the latter is broken by a partial polarization, the pairing order parameter is fundamentally changed to lead to a PDW state. Along this line, a further investigation into the phase of the pairing order parameter by tuning the spin background~\cite{Sun2020} may reveal more underlying novel structure of a doped Mott insulator.

\textit{Acknowledgments.---} Stimulating discussions with Jia-Xin Zhang, Jing-Yu Zhao, and Zheng Zhu are acknowledged. This work is partially supported by MOST of China (Grant No. 2021YFA1402101).

\bibliography{ref-cond}

\begin{thebibliography}{38}%
\makeatletter
\providecommand \@ifxundefined [1]{%
 \@ifx{#1\undefined}
}%
\providecommand \@ifnum [1]{%
 \ifnum #1\expandafter \@firstoftwo
 \else \expandafter \@secondoftwo
 \fi
}%
\providecommand \@ifx [1]{%
 \ifx #1\expandafter \@firstoftwo
 \else \expandafter \@secondoftwo
 \fi
}%
\providecommand \natexlab [1]{#1}%
\providecommand \enquote  [1]{``#1''}%
\providecommand \bibnamefont  [1]{#1}%
\providecommand \bibfnamefont [1]{#1}%
\providecommand \citenamefont [1]{#1}%
\providecommand \href@noop [0]{\@secondoftwo}%
\providecommand \href [0]{\begingroup \@sanitize@url \@href}%
\providecommand \@href[1]{\@@startlink{#1}\@@href}%
\providecommand \@@href[1]{\endgroup#1\@@endlink}%
\providecommand \@sanitize@url [0]{\catcode `\\12\catcode `\$12\catcode
  `\&12\catcode `\#12\catcode `\^12\catcode `\_12\catcode `\%12\relax}%
\providecommand \@@startlink[1]{}%
\providecommand \@@endlink[0]{}%
\providecommand \url  [0]{\begingroup\@sanitize@url \@url }%
\providecommand \@url [1]{\endgroup\@href {#1}{\urlprefix }}%
\providecommand \urlprefix  [0]{URL }%
\providecommand \Eprint [0]{\href }%
\providecommand \doibase [0]{http://dx.doi.org/}%
\providecommand \selectlanguage [0]{\@gobble}%
\providecommand \bibinfo  [0]{\@secondoftwo}%
\providecommand \bibfield  [0]{\@secondoftwo}%
\providecommand \translation [1]{[#1]}%
\providecommand \BibitemOpen [0]{}%
\providecommand \bibitemStop [0]{}%
\providecommand \bibitemNoStop [0]{.\EOS\space}%
\providecommand \EOS [0]{\spacefactor3000\relax}%
\providecommand \BibitemShut  [1]{\csname bibitem#1\endcsname}%
\let\auto@bib@innerbib\@empty
\bibitem [{\citenamefont {Agterberg}\ \emph {et~al.}(2020)\citenamefont
  {Agterberg}, \citenamefont {Davis}, \citenamefont {Edkins}, \citenamefont
  {Fradkin}, \citenamefont {{Van Harlingen}}, \citenamefont {Kivelson},
  \citenamefont {Lee}, \citenamefont {Radzihovsky}, \citenamefont {Tranquada},\
  and\ \citenamefont {Wang}}]{Agterberg2020}%
  \BibitemOpen
  \bibfield  {author} {\bibinfo {author} {\bibfnamefont {Daniel~F}\
  \bibnamefont {Agterberg}}, \bibinfo {author} {\bibfnamefont
  {J.~C.S{\'{e}}amus}\ \bibnamefont {Davis}}, \bibinfo {author} {\bibfnamefont
  {Stephen~D}\ \bibnamefont {Edkins}}, \bibinfo {author} {\bibfnamefont
  {Eduardo}\ \bibnamefont {Fradkin}}, \bibinfo {author} {\bibfnamefont
  {Dale~J.}\ \bibnamefont {{Van Harlingen}}}, \bibinfo {author} {\bibfnamefont
  {Steven~A}\ \bibnamefont {Kivelson}}, \bibinfo {author} {\bibfnamefont
  {Patrick~A}\ \bibnamefont {Lee}}, \bibinfo {author} {\bibfnamefont {Leo}\
  \bibnamefont {Radzihovsky}}, \bibinfo {author} {\bibfnamefont {John~M}\
  \bibnamefont {Tranquada}}, \ and\ \bibinfo {author} {\bibfnamefont {Yuxuan}\
  \bibnamefont {Wang}},\ }\href {\doibase
  10.1146/annurev-conmatphys-031119-050711} {\enquote {\bibinfo {title} {{The
  Physics of Pair-Density Waves: Cuprate Superconductors and beyond}},}\ }
  (\bibinfo {year} {2020})\BibitemShut {NoStop}%
\bibitem [{\citenamefont {Himeda}\ \emph {et~al.}(2002)\citenamefont {Himeda},
  \citenamefont {Kato},\ and\ \citenamefont {Ogata}}]{Himeda2002}%
  \BibitemOpen
  \bibfield  {author} {\bibinfo {author} {\bibfnamefont {A.}~\bibnamefont
  {Himeda}}, \bibinfo {author} {\bibfnamefont {T.}~\bibnamefont {Kato}}, \ and\
  \bibinfo {author} {\bibfnamefont {M.}~\bibnamefont {Ogata}},\ }\bibfield
  {title} {\enquote {\bibinfo {title} {{Stripe States with Spatially
  Oscillating $\textit{d}$-Wave Superconductivity in the Two-Dimensional
  $t$-$t'$-$J$ Model}},}\ }\href {\doibase 10.1103/PhysRevLett.88.117001}
  {\bibfield  {journal} {\bibinfo  {journal} {Physical Review Letters}\
  }\textbf {\bibinfo {volume} {88}},\ \bibinfo {pages} {117001} (\bibinfo
  {year} {2002})}\BibitemShut {NoStop}%
\bibitem [{\citenamefont {Li}\ \emph {et~al.}(2007)\citenamefont {Li},
  \citenamefont {H{\"{u}}cker}, \citenamefont {Gu}, \citenamefont {Tsvelik},\
  and\ \citenamefont {Tranquada}}]{Li2007}%
  \BibitemOpen
  \bibfield  {author} {\bibinfo {author} {\bibfnamefont {Q.}~\bibnamefont
  {Li}}, \bibinfo {author} {\bibfnamefont {M.}~\bibnamefont {H{\"{u}}cker}},
  \bibinfo {author} {\bibfnamefont {G~D}\ \bibnamefont {Gu}}, \bibinfo {author}
  {\bibfnamefont {A~M}\ \bibnamefont {Tsvelik}}, \ and\ \bibinfo {author}
  {\bibfnamefont {J~M}\ \bibnamefont {Tranquada}},\ }\bibfield  {title}
  {\enquote {\bibinfo {title} {{Two-dimensional superconducting fluctuations in
  stripe-ordered $\rm La_{1.875}Ba_{0.125}CuO_{4}$}},}\ }\href {\doibase
  10.1103/PhysRevLett.99.067001} {\bibfield  {journal} {\bibinfo  {journal}
  {Physical Review Letters}\ }\textbf {\bibinfo {volume} {99}},\ \bibinfo
  {pages} {4--7} (\bibinfo {year} {2007})},\ \Eprint
  {http://arxiv.org/abs/0703357} {arXiv:0703357 [cond-mat]} \BibitemShut
  {NoStop}%
\bibitem [{\citenamefont {Berg}\ \emph {et~al.}(2007)\citenamefont {Berg},
  \citenamefont {Fradkin}, \citenamefont {Kim}, \citenamefont {Kivelson},
  \citenamefont {Oganesyan}, \citenamefont {Tranquada},\ and\ \citenamefont
  {Zhang}}]{Berg2007}%
  \BibitemOpen
  \bibfield  {author} {\bibinfo {author} {\bibfnamefont {E.}~\bibnamefont
  {Berg}}, \bibinfo {author} {\bibfnamefont {E.}~\bibnamefont {Fradkin}},
  \bibinfo {author} {\bibfnamefont {E.-A.}\ \bibnamefont {Kim}}, \bibinfo
  {author} {\bibfnamefont {S.~A.}\ \bibnamefont {Kivelson}}, \bibinfo {author}
  {\bibfnamefont {V.}~\bibnamefont {Oganesyan}}, \bibinfo {author}
  {\bibfnamefont {J.~M.}\ \bibnamefont {Tranquada}}, \ and\ \bibinfo {author}
  {\bibfnamefont {S.~C.}\ \bibnamefont {Zhang}},\ }\bibfield  {title} {\enquote
  {\bibinfo {title} {{Dynamical Layer Decoupling in a Stripe-Ordered High-$T_c$
  Superconductor}},}\ }\href {\doibase 10.1103/PhysRevLett.99.127003}
  {\bibfield  {journal} {\bibinfo  {journal} {Physical Review Letters}\
  }\textbf {\bibinfo {volume} {99}},\ \bibinfo {pages} {127003} (\bibinfo
  {year} {2007})},\ \Eprint {http://arxiv.org/abs/0704.1240} {arXiv:0704.1240}
  \BibitemShut {NoStop}%
\bibitem [{\citenamefont {Lozano}\ \emph {et~al.}(2022)\citenamefont {Lozano},
  \citenamefont {Ren}, \citenamefont {Gu}, \citenamefont {Tsvelik},
  \citenamefont {Tranquada},\ and\ \citenamefont {Li}}]{Lozano2022}%
  \BibitemOpen
  \bibfield  {author} {\bibinfo {author} {\bibfnamefont {P.~M.}\ \bibnamefont
  {Lozano}}, \bibinfo {author} {\bibfnamefont {Tianhao}\ \bibnamefont {Ren}},
  \bibinfo {author} {\bibfnamefont {G.~D.}\ \bibnamefont {Gu}}, \bibinfo
  {author} {\bibfnamefont {A.~M.}\ \bibnamefont {Tsvelik}}, \bibinfo {author}
  {\bibfnamefont {J.~M.}\ \bibnamefont {Tranquada}}, \ and\ \bibinfo {author}
  {\bibfnamefont {Qiang}\ \bibnamefont {Li}},\ }\bibfield  {title} {\enquote
  {\bibinfo {title} {{Testing for pair density wave order in $\rm
  La_{1.875}Ba_{0.125}CuO_4$}},}\ }\href {\doibase 10.1103/PhysRevB.106.174510}
  {\bibfield  {journal} {\bibinfo  {journal} {Physical Review B}\ }\textbf
  {\bibinfo {volume} {106}},\ \bibinfo {pages} {174510} (\bibinfo {year}
  {2022})},\ \Eprint {http://arxiv.org/abs/2110.05513} {arXiv:2110.05513}
  \BibitemShut {NoStop}%
\bibitem [{\citenamefont {Hamidian}\ \emph {et~al.}(2016)\citenamefont
  {Hamidian}, \citenamefont {Edkins}, \citenamefont {Joo}, \citenamefont
  {Kostin}, \citenamefont {Eisaki}, \citenamefont {Uchida}, \citenamefont
  {Lawler}, \citenamefont {Kim}, \citenamefont {Mackenzie}, \citenamefont
  {Fujita}, \citenamefont {Lee},\ and\ \citenamefont {Davis}}]{Hamidian2016}%
  \BibitemOpen
  \bibfield  {author} {\bibinfo {author} {\bibfnamefont {M.~H.}\ \bibnamefont
  {Hamidian}}, \bibinfo {author} {\bibfnamefont {S.~D.}\ \bibnamefont
  {Edkins}}, \bibinfo {author} {\bibfnamefont {Sang~Hyun}\ \bibnamefont {Joo}},
  \bibinfo {author} {\bibfnamefont {A.}~\bibnamefont {Kostin}}, \bibinfo
  {author} {\bibfnamefont {H.}~\bibnamefont {Eisaki}}, \bibinfo {author}
  {\bibfnamefont {S.}~\bibnamefont {Uchida}}, \bibinfo {author} {\bibfnamefont
  {M.~J.}\ \bibnamefont {Lawler}}, \bibinfo {author} {\bibfnamefont {E.-A.}\
  \bibnamefont {Kim}}, \bibinfo {author} {\bibfnamefont {A.~P.}\ \bibnamefont
  {Mackenzie}}, \bibinfo {author} {\bibfnamefont {K.}~\bibnamefont {Fujita}},
  \bibinfo {author} {\bibfnamefont {Jinho}\ \bibnamefont {Lee}}, \ and\
  \bibinfo {author} {\bibfnamefont {J.~C.~S{\'{e}}amus}\ \bibnamefont
  {Davis}},\ }\bibfield  {title} {\enquote {\bibinfo {title} {{Detection of a
  Cooper-pair density wave in $\rm Bi_2Sr_2CaCu_2O_{8+x}$}},}\ }\href {\doibase
  10.1038/nature17411} {\bibfield  {journal} {\bibinfo  {journal} {Nature}\
  }\textbf {\bibinfo {volume} {532}},\ \bibinfo {pages} {343--347} (\bibinfo
  {year} {2016})},\ \Eprint {http://arxiv.org/abs/1511.08124}
  {arXiv:1511.08124} \BibitemShut {NoStop}%
\bibitem [{\citenamefont {Ruan}\ \emph {et~al.}(2018)\citenamefont {Ruan},
  \citenamefont {Li}, \citenamefont {Hu}, \citenamefont {Hao}, \citenamefont
  {Li}, \citenamefont {Cai}, \citenamefont {Zhou}, \citenamefont {Lee},\ and\
  \citenamefont {Wang}}]{Ruan2018}%
  \BibitemOpen
  \bibfield  {author} {\bibinfo {author} {\bibfnamefont {Wei}\ \bibnamefont
  {Ruan}}, \bibinfo {author} {\bibfnamefont {Xintong}\ \bibnamefont {Li}},
  \bibinfo {author} {\bibfnamefont {Cheng}\ \bibnamefont {Hu}}, \bibinfo
  {author} {\bibfnamefont {Zhenqi}\ \bibnamefont {Hao}}, \bibinfo {author}
  {\bibfnamefont {Haiwei}\ \bibnamefont {Li}}, \bibinfo {author} {\bibfnamefont
  {Peng}\ \bibnamefont {Cai}}, \bibinfo {author} {\bibfnamefont {Xingjiang}\
  \bibnamefont {Zhou}}, \bibinfo {author} {\bibfnamefont {Dung~Hai}\
  \bibnamefont {Lee}}, \ and\ \bibinfo {author} {\bibfnamefont {Yayu}\
  \bibnamefont {Wang}},\ }\href {\doibase 10.1038/s41567-018-0276-8} {\enquote
  {\bibinfo {title} {{Visualization of the periodic modulation of Cooper
  pairing in a cuprate superconductor}},}\ } (\bibinfo {year}
  {2018})\BibitemShut {NoStop}%
\bibitem [{\citenamefont {Edkins}\ \emph {et~al.}(2019)\citenamefont {Edkins},
  \citenamefont {Kostin}, \citenamefont {Fujita}, \citenamefont {Mackenzie},
  \citenamefont {Eisaki}, \citenamefont {Uchida}, \citenamefont {Sachdev},
  \citenamefont {Lawler}, \citenamefont {Kim}, \citenamefont {Davis},\ and\
  \citenamefont {Hamidian}}]{Edkins2019}%
  \BibitemOpen
  \bibfield  {author} {\bibinfo {author} {\bibfnamefont {S~D}\ \bibnamefont
  {Edkins}}, \bibinfo {author} {\bibfnamefont {A}~\bibnamefont {Kostin}},
  \bibinfo {author} {\bibfnamefont {K}~\bibnamefont {Fujita}}, \bibinfo
  {author} {\bibfnamefont {A~P}\ \bibnamefont {Mackenzie}}, \bibinfo {author}
  {\bibfnamefont {H}~\bibnamefont {Eisaki}}, \bibinfo {author} {\bibfnamefont
  {S}~\bibnamefont {Uchida}}, \bibinfo {author} {\bibfnamefont {Subir}\
  \bibnamefont {Sachdev}}, \bibinfo {author} {\bibfnamefont {Michael~J}\
  \bibnamefont {Lawler}}, \bibinfo {author} {\bibfnamefont {E.~A.}\
  \bibnamefont {Kim}}, \bibinfo {author} {\bibfnamefont {J.~C.S{\'{e}}amus}\
  \bibnamefont {Davis}}, \ and\ \bibinfo {author} {\bibfnamefont {M~H}\
  \bibnamefont {Hamidian}},\ }\bibfield  {title} {\enquote {\bibinfo {title}
  {{Magnetic field-induced pair density wave state in the cuprate vortex
  halo}},}\ }\href {\doibase 10.1126/science.aat1773} {\bibfield  {journal}
  {\bibinfo  {journal} {Science}\ }\textbf {\bibinfo {volume} {364}},\ \bibinfo
  {pages} {976--980} (\bibinfo {year} {2019})},\ \Eprint
  {http://arxiv.org/abs/1802.04673} {arXiv:1802.04673} \BibitemShut {NoStop}%
\bibitem [{\citenamefont {Fulde}\ and\ \citenamefont
  {Ferrell}(1964)}]{Fulde1964}%
  \BibitemOpen
  \bibfield  {author} {\bibinfo {author} {\bibfnamefont {Peter}\ \bibnamefont
  {Fulde}}\ and\ \bibinfo {author} {\bibfnamefont {Richard~A.}\ \bibnamefont
  {Ferrell}},\ }\bibfield  {title} {\enquote {\bibinfo {title}
  {{Superconductivity in a Strong Spin-Exchange Field}},}\ }\href {\doibase
  10.1103/PhysRev.135.A550} {\bibfield  {journal} {\bibinfo  {journal}
  {Physical Review}\ }\textbf {\bibinfo {volume} {135}},\ \bibinfo {pages}
  {A550--A563} (\bibinfo {year} {1964})}\BibitemShut {NoStop}%
\bibitem [{\citenamefont {Larkin}\ and\ \citenamefont
  {Ovchinnikov}(1965)}]{Larkin1965}%
  \BibitemOpen
  \bibfield  {author} {\bibinfo {author} {\bibfnamefont {A.~I.}\ \bibnamefont
  {Larkin}}\ and\ \bibinfo {author} {\bibfnamefont {Y.~N.}\ \bibnamefont
  {Ovchinnikov}},\ }\bibfield  {title} {\enquote {\bibinfo {title} {{Nonuniform
  state of superconductors}},}\ }\href
  {https://www.osti.gov/biblio/4653415-nonuniform-state-superconductors}
  {\bibfield  {journal} {\bibinfo  {journal} {Soviet Physics-JETP}\ }\textbf
  {\bibinfo {volume} {20}},\ \bibinfo {pages} {762--770} (\bibinfo {year}
  {1965})}\BibitemShut {NoStop}%
\bibitem [{\citenamefont {Berg}\ \emph {et~al.}(2009)\citenamefont {Berg},
  \citenamefont {Fradkin}, \citenamefont {Kivelson},\ and\ \citenamefont
  {Tranquada}}]{Berg2009}%
  \BibitemOpen
  \bibfield  {author} {\bibinfo {author} {\bibfnamefont {Erez}\ \bibnamefont
  {Berg}}, \bibinfo {author} {\bibfnamefont {Eduardo}\ \bibnamefont {Fradkin}},
  \bibinfo {author} {\bibfnamefont {Steven~A.}\ \bibnamefont {Kivelson}}, \
  and\ \bibinfo {author} {\bibfnamefont {John~M.}\ \bibnamefont {Tranquada}},\
  }\bibfield  {title} {\enquote {\bibinfo {title} {{Striped superconductors:
  how spin, charge and superconducting orders intertwine in the cuprates}},}\
  }\href {\doibase 10.1088/1367-2630/11/11/115004} {\bibfield  {journal}
  {\bibinfo  {journal} {New Journal of Physics}\ }\textbf {\bibinfo {volume}
  {11}},\ \bibinfo {pages} {115004} (\bibinfo {year} {2009})}\BibitemShut
  {NoStop}%
\bibitem [{\citenamefont {Berg}\ \emph {et~al.}(2010)\citenamefont {Berg},
  \citenamefont {Fradkin},\ and\ \citenamefont {Kivelson}}]{Berg2010}%
  \BibitemOpen
  \bibfield  {author} {\bibinfo {author} {\bibfnamefont {Erez}\ \bibnamefont
  {Berg}}, \bibinfo {author} {\bibfnamefont {Eduardo}\ \bibnamefont {Fradkin}},
  \ and\ \bibinfo {author} {\bibfnamefont {Steven~A.}\ \bibnamefont
  {Kivelson}},\ }\bibfield  {title} {\enquote {\bibinfo {title}
  {{Pair-density-wave correlations in the Kondo-Heisenberg model}},}\ }\href
  {\doibase 10.1103/PhysRevLett.105.146403} {\bibfield  {journal} {\bibinfo
  {journal} {Physical Review Letters}\ }\textbf {\bibinfo {volume} {105}},\
  \bibinfo {pages} {2--5} (\bibinfo {year} {2010})}\BibitemShut {NoStop}%
\bibitem [{\citenamefont {Loder}\ \emph {et~al.}(2010)\citenamefont {Loder},
  \citenamefont {Kampf},\ and\ \citenamefont {Kopp}}]{Loder2010}%
  \BibitemOpen
  \bibfield  {author} {\bibinfo {author} {\bibfnamefont {Florian}\ \bibnamefont
  {Loder}}, \bibinfo {author} {\bibfnamefont {Arno~P.}\ \bibnamefont {Kampf}},
  \ and\ \bibinfo {author} {\bibfnamefont {Thilo}\ \bibnamefont {Kopp}},\
  }\bibfield  {title} {\enquote {\bibinfo {title} {{Superconducting state with
  a finite-momentum pairing mechanism in zero external magnetic field}},}\
  }\href {\doibase 10.1103/PhysRevB.81.020511} {\bibfield  {journal} {\bibinfo
  {journal} {Physical Review B}\ }\textbf {\bibinfo {volume} {81}},\ \bibinfo
  {pages} {020511} (\bibinfo {year} {2010})}\BibitemShut {NoStop}%
\bibitem [{\citenamefont {Loder}\ \emph {et~al.}(2011)\citenamefont {Loder},
  \citenamefont {Graser}, \citenamefont {Kampf},\ and\ \citenamefont
  {Kopp}}]{Loder2011}%
  \BibitemOpen
  \bibfield  {author} {\bibinfo {author} {\bibfnamefont {Florian}\ \bibnamefont
  {Loder}}, \bibinfo {author} {\bibfnamefont {Siegfried}\ \bibnamefont
  {Graser}}, \bibinfo {author} {\bibfnamefont {Arno~P.}\ \bibnamefont {Kampf}},
  \ and\ \bibinfo {author} {\bibfnamefont {Thilo}\ \bibnamefont {Kopp}},\
  }\bibfield  {title} {\enquote {\bibinfo {title} {{Mean-field pairing theory
  for the charge-stripe phase of high-temperature cuprate superconductors}},}\
  }\href {\doibase 10.1103/PhysRevLett.107.187001} {\bibfield  {journal}
  {\bibinfo  {journal} {Physical Review Letters}\ }\textbf {\bibinfo {volume}
  {107}},\ \bibinfo {pages} {1--4} (\bibinfo {year} {2011})}\BibitemShut
  {NoStop}%
\bibitem [{\citenamefont {Lee}(2014)}]{Lee2014}%
  \BibitemOpen
  \bibfield  {author} {\bibinfo {author} {\bibfnamefont {Patrick~A.}\
  \bibnamefont {Lee}},\ }\bibfield  {title} {\enquote {\bibinfo {title}
  {{Amperean Pairing and the Pseudogap Phase of Cuprate Superconductors}},}\
  }\href {\doibase 10.1103/PhysRevX.4.031017} {\bibfield  {journal} {\bibinfo
  {journal} {Physical Review X}\ }\textbf {\bibinfo {volume} {4}},\ \bibinfo
  {pages} {031017} (\bibinfo {year} {2014})},\ \Eprint
  {http://arxiv.org/abs/1401.0519} {arXiv:1401.0519} \BibitemShut {NoStop}%
\bibitem [{\citenamefont {W{\aa}rdh}\ and\ \citenamefont
  {Granath}(2017)}]{Wardh2017}%
  \BibitemOpen
  \bibfield  {author} {\bibinfo {author} {\bibfnamefont {Jonatan}\ \bibnamefont
  {W{\aa}rdh}}\ and\ \bibinfo {author} {\bibfnamefont {Mats}\ \bibnamefont
  {Granath}},\ }\bibfield  {title} {\enquote {\bibinfo {title} {{Effective
  model for a supercurrent in a pair-density wave}},}\ }\href {\doibase
  10.1103/PhysRevB.96.224503} {\bibfield  {journal} {\bibinfo  {journal}
  {Physical Review B}\ }\textbf {\bibinfo {volume} {96}},\ \bibinfo {pages}
  {1--11} (\bibinfo {year} {2017})},\ \Eprint {http://arxiv.org/abs/1703.03781}
  {arXiv:1703.03781} \BibitemShut {NoStop}%
\bibitem [{\citenamefont {W{\aa}rdh}\ \emph {et~al.}(2018)\citenamefont
  {W{\aa}rdh}, \citenamefont {Andersen},\ and\ \citenamefont
  {Granath}}]{Wardh2018}%
  \BibitemOpen
  \bibfield  {author} {\bibinfo {author} {\bibfnamefont {Jonatan}\ \bibnamefont
  {W{\aa}rdh}}, \bibinfo {author} {\bibfnamefont {Brian~M.}\ \bibnamefont
  {Andersen}}, \ and\ \bibinfo {author} {\bibfnamefont {Mats}\ \bibnamefont
  {Granath}},\ }\bibfield  {title} {\enquote {\bibinfo {title} {{Suppression of
  superfluid stiffness near a Lifshitz-point instability to finite-momentum
  superconductivity}},}\ }\href {\doibase 10.1103/PhysRevB.98.224501}
  {\bibfield  {journal} {\bibinfo  {journal} {Physical Review B}\ }\textbf
  {\bibinfo {volume} {98}},\ \bibinfo {pages} {1--10} (\bibinfo {year}
  {2018})},\ \Eprint {http://arxiv.org/abs/1807.05303} {arXiv:1807.05303}
  \BibitemShut {NoStop}%
\bibitem [{\citenamefont {Setty}\ \emph {et~al.}(2021)\citenamefont {Setty},
  \citenamefont {Fanfarillo},\ and\ \citenamefont {Hirschfeld}}]{Setty2021}%
  \BibitemOpen
  \bibfield  {author} {\bibinfo {author} {\bibfnamefont {Chandan}\ \bibnamefont
  {Setty}}, \bibinfo {author} {\bibfnamefont {Laura}\ \bibnamefont
  {Fanfarillo}}, \ and\ \bibinfo {author} {\bibfnamefont {P.~J.}\ \bibnamefont
  {Hirschfeld}},\ }\bibfield  {title} {\enquote {\bibinfo {title} {{Microscopic
  mechanism for fluctuating pair density wave}},}\ }\href
  {http://arxiv.org/abs/2110.13138} {\ ,\ \bibinfo {pages} {1--13} (\bibinfo
  {year} {2021})},\ \Eprint {http://arxiv.org/abs/2110.13138}
  {arXiv:2110.13138} \BibitemShut {NoStop}%
\bibitem [{\citenamefont {Setty}\ \emph {et~al.}(2022)\citenamefont {Setty},
  \citenamefont {Zhao}, \citenamefont {Fanfarillo}, \citenamefont {Huang},
  \citenamefont {Hirschfeld}, \citenamefont {Phillips},\ and\ \citenamefont
  {Yang}}]{Setty2022}%
  \BibitemOpen
  \bibfield  {author} {\bibinfo {author} {\bibfnamefont {Chandan}\ \bibnamefont
  {Setty}}, \bibinfo {author} {\bibfnamefont {Jinchao}\ \bibnamefont {Zhao}},
  \bibinfo {author} {\bibfnamefont {Laura}\ \bibnamefont {Fanfarillo}},
  \bibinfo {author} {\bibfnamefont {Edwin~W.}\ \bibnamefont {Huang}}, \bibinfo
  {author} {\bibfnamefont {Peter~J.}\ \bibnamefont {Hirschfeld}}, \bibinfo
  {author} {\bibfnamefont {Philip~W.}\ \bibnamefont {Phillips}}, \ and\
  \bibinfo {author} {\bibfnamefont {Kun}\ \bibnamefont {Yang}},\ }\bibfield
  {title} {\enquote {\bibinfo {title} {{Exact solution for finite
  center-of-mass momentum Cooper pairing}},}\ }\href
  {http://arxiv.org/abs/2209.10568} {\  (\bibinfo {year} {2022})},\ \Eprint
  {http://arxiv.org/abs/2209.10568} {arXiv:2209.10568} \BibitemShut {NoStop}%
\bibitem [{\citenamefont {Jiang}(2022)}]{Jiang2022}%
  \BibitemOpen
  \bibfield  {author} {\bibinfo {author} {\bibfnamefont {Hong-Chen}\
  \bibnamefont {Jiang}},\ }\bibfield  {title} {\enquote {\bibinfo {title}
  {{Pair density wave in doped three-band Hubbard model on square lattice}},}\
  }\href {http://arxiv.org/abs/2209.11381} {\  (\bibinfo {year} {2022})},\
  \Eprint {http://arxiv.org/abs/2209.11381} {arXiv:2209.11381} \BibitemShut
  {NoStop}%
\bibitem [{\citenamefont {Wu}\ \emph {et~al.}(2022)\citenamefont {Wu},
  \citenamefont {Nosov}, \citenamefont {Patel},\ and\ \citenamefont
  {Raghu}}]{Wu2022}%
  \BibitemOpen
  \bibfield  {author} {\bibinfo {author} {\bibfnamefont {Yi-Ming}\ \bibnamefont
  {Wu}}, \bibinfo {author} {\bibfnamefont {P.~A.}\ \bibnamefont {Nosov}},
  \bibinfo {author} {\bibfnamefont {Aavishkar~A.}\ \bibnamefont {Patel}}, \
  and\ \bibinfo {author} {\bibfnamefont {S.}~\bibnamefont {Raghu}},\ }\bibfield
   {title} {\enquote {\bibinfo {title} {{Pair density wave order from electron
  repulsion}},}\ }\href {http://arxiv.org/abs/2209.09254} {\  (\bibinfo {year}
  {2022})},\ \Eprint {http://arxiv.org/abs/2209.09254} {arXiv:2209.09254}
  \BibitemShut {NoStop}%
\bibitem [{\citenamefont {Poilblanc}\ \emph {et~al.}(1995)\citenamefont
  {Poilblanc}, \citenamefont {Scalapino},\ and\ \citenamefont
  {Hanke}}]{Poilblanc1995}%
  \BibitemOpen
  \bibfield  {author} {\bibinfo {author} {\bibfnamefont {D.}~\bibnamefont
  {Poilblanc}}, \bibinfo {author} {\bibfnamefont {D.~J.}\ \bibnamefont
  {Scalapino}}, \ and\ \bibinfo {author} {\bibfnamefont {W.}~\bibnamefont
  {Hanke}},\ }\bibfield  {title} {\enquote {\bibinfo {title} {{Spin and charge
  modes of the t - J ladder}},}\ }\href {\doibase 10.1103/PhysRevB.52.6796}
  {\bibfield  {journal} {\bibinfo  {journal} {Physical Review B}\ }\textbf
  {\bibinfo {volume} {52}},\ \bibinfo {pages} {6796--6800} (\bibinfo {year}
  {1995})}\BibitemShut {NoStop}%
\bibitem [{\citenamefont {Roux}\ \emph {et~al.}(2006)\citenamefont {Roux},
  \citenamefont {White}, \citenamefont {Capponi},\ and\ \citenamefont
  {Poilblanc}}]{Roux2006}%
  \BibitemOpen
  \bibfield  {author} {\bibinfo {author} {\bibfnamefont {G.}~\bibnamefont
  {Roux}}, \bibinfo {author} {\bibfnamefont {S.~R.}\ \bibnamefont {White}},
  \bibinfo {author} {\bibfnamefont {S.}~\bibnamefont {Capponi}}, \ and\
  \bibinfo {author} {\bibfnamefont {D.}~\bibnamefont {Poilblanc}},\ }\bibfield
  {title} {\enquote {\bibinfo {title} {{Zeeman effect in superconducting
  two-leg ladders: Irrational magnetization plateaus and exceeding the Pauli
  limit}},}\ }\href {\doibase 10.1103/PhysRevLett.97.087207} {\bibfield
  {journal} {\bibinfo  {journal} {Physical Review Letters}\ }\textbf {\bibinfo
  {volume} {97}},\ \bibinfo {pages} {1--4} (\bibinfo {year} {2006})},\ \Eprint
  {http://arxiv.org/abs/0512025} {arXiv:0512025 [cond-mat]} \BibitemShut
  {NoStop}%
\bibitem [{\citenamefont {Jiang}\ \emph {et~al.}(2020)\citenamefont {Jiang},
  \citenamefont {Chen},\ and\ \citenamefont {Weng}}]{Jiang2020}%
  \BibitemOpen
  \bibfield  {author} {\bibinfo {author} {\bibfnamefont {Hong-Chen}\
  \bibnamefont {Jiang}}, \bibinfo {author} {\bibfnamefont {Shuai}\ \bibnamefont
  {Chen}}, \ and\ \bibinfo {author} {\bibfnamefont {Zheng-Yu}\ \bibnamefont
  {Weng}},\ }\bibfield  {title} {\enquote {\bibinfo {title} {{Critical role of
  the sign structure in the doped Mott insulator: Luther-Emery versus
  Fermi-liquid-like state in quasi-one-dimensional ladders}},}\ }\href
  {\doibase 10.1103/PhysRevB.102.104512} {\bibfield  {journal} {\bibinfo
  {journal} {Physical Review B}\ }\textbf {\bibinfo {volume} {102}},\ \bibinfo
  {pages} {104512} (\bibinfo {year} {2020})}\BibitemShut {NoStop}%
\bibitem [{\citenamefont {Sun}\ \emph {et~al.}(2020)\citenamefont {Sun},
  \citenamefont {Zhu},\ and\ \citenamefont {Weng}}]{Sun2020}%
  \BibitemOpen
  \bibfield  {author} {\bibinfo {author} {\bibfnamefont {Rong-Yang}\
  \bibnamefont {Sun}}, \bibinfo {author} {\bibfnamefont {Zheng}\ \bibnamefont
  {Zhu}}, \ and\ \bibinfo {author} {\bibfnamefont {Zheng-Yu}\ \bibnamefont
  {Weng}},\ }\bibfield  {title} {\enquote {\bibinfo {title} {{Complex phase
  diagram of doped XXZ ladder: Localization and pairing}},}\ }\href {\doibase
  10.1103/PhysRevResearch.2.033007} {\bibfield  {journal} {\bibinfo  {journal}
  {Physical Review Research}\ }\textbf {\bibinfo {volume} {2}},\ \bibinfo
  {pages} {033007} (\bibinfo {year} {2020})},\ \Eprint
  {http://arxiv.org/abs/2002.03529} {arXiv:2002.03529} \BibitemShut {NoStop}%
\bibitem [{\citenamefont {Shinjo}\ \emph {et~al.}(2021)\citenamefont {Shinjo},
  \citenamefont {Sota},\ and\ \citenamefont {Tohyama}}]{Shinjo2021}%
  \BibitemOpen
  \bibfield  {author} {\bibinfo {author} {\bibfnamefont {Kazuya}\ \bibnamefont
  {Shinjo}}, \bibinfo {author} {\bibfnamefont {Shigetoshi}\ \bibnamefont
  {Sota}}, \ and\ \bibinfo {author} {\bibfnamefont {Takami}\ \bibnamefont
  {Tohyama}},\ }\bibfield  {title} {\enquote {\bibinfo {title} {{Effect of
  phase string on single-hole dynamics in the two-leg Hubbard ladder}},}\
  }\href {\doibase 10.1103/PhysRevB.103.035141} {\bibfield  {journal} {\bibinfo
   {journal} {Physical Review B}\ }\textbf {\bibinfo {volume} {103}},\ \bibinfo
  {pages} {035141} (\bibinfo {year} {2021})},\ \Eprint
  {http://arxiv.org/abs/2011.10686} {arXiv:2011.10686} \BibitemShut {NoStop}%
\bibitem [{\citenamefont {Luther}\ and\ \citenamefont
  {Emery}(1974)}]{Luther1974}%
  \BibitemOpen
  \bibfield  {author} {\bibinfo {author} {\bibfnamefont {A.}~\bibnamefont
  {Luther}}\ and\ \bibinfo {author} {\bibfnamefont {V.~J.}\ \bibnamefont
  {Emery}},\ }\bibfield  {title} {\enquote {\bibinfo {title} {{Backward
  Scattering in the One-Dimensional Electron Gas}},}\ }\href {\doibase
  10.1103/PhysRevLett.33.589} {\bibfield  {journal} {\bibinfo  {journal}
  {Physical Review Letters}\ }\textbf {\bibinfo {volume} {33}} (\bibinfo {year}
  {1974}),\ 10.1103/PhysRevLett.33.589}\BibitemShut {NoStop}%
\bibitem [{\citenamefont {Zhu}\ \emph {et~al.}(2018)\citenamefont {Zhu},
  \citenamefont {Sheng},\ and\ \citenamefont {Weng}}]{Zhu2018}%
  \BibitemOpen
  \bibfield  {author} {\bibinfo {author} {\bibfnamefont {Zheng}\ \bibnamefont
  {Zhu}}, \bibinfo {author} {\bibfnamefont {D~N}\ \bibnamefont {Sheng}}, \ and\
  \bibinfo {author} {\bibfnamefont {Zheng~Yu}\ \bibnamefont {Weng}},\
  }\bibfield  {title} {\enquote {\bibinfo {title} {{Pairing versus phase
  coherence of doped holes in distinct quantum spin backgrounds}},}\ }\href
  {\doibase 10.1103/PhysRevB.97.115144} {\bibfield  {journal} {\bibinfo
  {journal} {Physical Review B}\ }\textbf {\bibinfo {volume} {97}},\ \bibinfo
  {pages} {1--8} (\bibinfo {year} {2018})},\ \Eprint
  {http://arxiv.org/abs/1706.02305} {arXiv:1706.02305} \BibitemShut {NoStop}%
\bibitem [{\citenamefont {Chen}\ \emph {et~al.}(2018)\citenamefont {Chen},
  \citenamefont {Zhu},\ and\ \citenamefont {Weng}}]{Chen2018a}%
  \BibitemOpen
  \bibfield  {author} {\bibinfo {author} {\bibfnamefont {Shuai}\ \bibnamefont
  {Chen}}, \bibinfo {author} {\bibfnamefont {Zheng}\ \bibnamefont {Zhu}}, \
  and\ \bibinfo {author} {\bibfnamefont {Zheng-Yu}\ \bibnamefont {Weng}},\
  }\bibfield  {title} {\enquote {\bibinfo {title} {{Two-hole ground state
  wavefunction: Non-BCS pairing in a t-J two-leg ladder}},}\ }\href {\doibase
  10.1103/PhysRevB.98.245138} {\bibfield  {journal} {\bibinfo  {journal}
  {Physical Review B}\ }\textbf {\bibinfo {volume} {98}},\ \bibinfo {pages}
  {245138} (\bibinfo {year} {2018})},\ \Eprint
  {http://arxiv.org/abs/1808.06173} {arXiv:1808.06173} \BibitemShut {NoStop}%
\bibitem [{\citenamefont {White}(1992)}]{White1992}%
  \BibitemOpen
  \bibfield  {author} {\bibinfo {author} {\bibfnamefont {Steven~R.}\
  \bibnamefont {White}},\ }\bibfield  {title} {\enquote {\bibinfo {title}
  {{Density matrix formulation for quantum renormalization groups}},}\ }\href
  {\doibase 10.1103/PhysRevLett.69.2863} {\bibfield  {journal} {\bibinfo
  {journal} {Physical Review Letters}\ }\textbf {\bibinfo {volume} {69}}
  (\bibinfo {year} {1992}),\ 10.1103/PhysRevLett.69.2863}\BibitemShut {NoStop}%
\bibitem [{SM()}]{SM}%
  \BibitemOpen
  \href@noop {} {}\bibinfo {howpublished} {See Supplemental Material for more
  data analyses supporting the phase diagram and the conclusion on the pairing
  amplitude and phase.}\BibitemShut {Stop}%
\bibitem [{\citenamefont {GraceQ}()}]{graceq}%
  \BibitemOpen
  \bibfield  {author} {\bibinfo {author} {\bibnamefont {GraceQ}},\ }\href@noop
  {} {}\bibinfo {howpublished}
  {\href{gracequantum.org}{www.gracequantum.org}}\BibitemShut {NoStop}%
\bibitem [{\citenamefont {Zhao}\ \emph {et~al.}(2022)\citenamefont {Zhao},
  \citenamefont {Chen}, \citenamefont {Zhang},\ and\ \citenamefont
  {Weng}}]{Zhao2022b}%
  \BibitemOpen
  \bibfield  {author} {\bibinfo {author} {\bibfnamefont {Jing~Yu}\ \bibnamefont
  {Zhao}}, \bibinfo {author} {\bibfnamefont {Shuai~A}\ \bibnamefont {Chen}},
  \bibinfo {author} {\bibfnamefont {Hao~Kai}\ \bibnamefont {Zhang}}, \ and\
  \bibinfo {author} {\bibfnamefont {Zheng~Yu}\ \bibnamefont {Weng}},\
  }\bibfield  {title} {\enquote {\bibinfo {title} {{Two-Hole Ground State:
  Dichotomy in Pairing Symmetry}},}\ }\href {\doibase
  10.1103/PhysRevX.12.011062} {\bibfield  {journal} {\bibinfo  {journal}
  {Physical Review X}\ }\textbf {\bibinfo {volume} {12}},\ \bibinfo {pages}
  {1--22} (\bibinfo {year} {2022})},\ \Eprint {http://arxiv.org/abs/2106.14898}
  {arXiv:2106.14898} \BibitemShut {NoStop}%
\bibitem [{\citenamefont {Wu}\ \emph {et~al.}(2008)\citenamefont {Wu},
  \citenamefont {Weng},\ and\ \citenamefont {Zaanen}}]{Wu2008}%
  \BibitemOpen
  \bibfield  {author} {\bibinfo {author} {\bibfnamefont {K}~\bibnamefont {Wu}},
  \bibinfo {author} {\bibfnamefont {Z~Y}\ \bibnamefont {Weng}}, \ and\ \bibinfo
  {author} {\bibfnamefont {J}~\bibnamefont {Zaanen}},\ }\bibfield  {title}
  {\enquote {\bibinfo {title} {{Sign structure of the t-J model}},}\ }\href
  {\doibase 10.1103/PhysRevB.77.155102} {\bibfield  {journal} {\bibinfo
  {journal} {Physical Review B - Condensed Matter and Materials Physics}\
  }\textbf {\bibinfo {volume} {77}},\ \bibinfo {pages} {1--5} (\bibinfo {year}
  {2008})}\BibitemShut {NoStop}%
\bibitem [{\citenamefont {Zhu}\ \emph {et~al.}(2013)\citenamefont {Zhu},
  \citenamefont {Jiang}, \citenamefont {Qi}, \citenamefont {Tian},\ and\
  \citenamefont {Weng}}]{Zhu2012}%
  \BibitemOpen
  \bibfield  {author} {\bibinfo {author} {\bibfnamefont {Zheng}\ \bibnamefont
  {Zhu}}, \bibinfo {author} {\bibfnamefont {Hong~Chen}\ \bibnamefont {Jiang}},
  \bibinfo {author} {\bibfnamefont {Yang}\ \bibnamefont {Qi}}, \bibinfo
  {author} {\bibfnamefont {Chu-Shun}\ \bibnamefont {Tian}}, \ and\ \bibinfo
  {author} {\bibfnamefont {Zheng~Yu}\ \bibnamefont {Weng}},\ }\bibfield
  {title} {\enquote {\bibinfo {title} {{Strong correlation induced charge
  localization in antiferromagnets}},}\ }\href {\doibase 10.1038/srep02586}
  {\bibfield  {journal} {\bibinfo  {journal} {Scientific Reports}\ }\textbf
  {\bibinfo {volume} {3}} (\bibinfo {year} {2013}),\ 10.1038/srep02586},\
  \Eprint {http://arxiv.org/abs/1212.6634} {arXiv:1212.6634} \BibitemShut
  {NoStop}%
\bibitem [{\citenamefont {Sheng}\ \emph {et~al.}(1996)\citenamefont {Sheng},
  \citenamefont {Chen},\ and\ \citenamefont {Weng}}]{Sheng1996}%
  \BibitemOpen
  \bibfield  {author} {\bibinfo {author} {\bibfnamefont {D~N}\ \bibnamefont
  {Sheng}}, \bibinfo {author} {\bibfnamefont {Y~C}\ \bibnamefont {Chen}}, \
  and\ \bibinfo {author} {\bibfnamefont {Z~Y}\ \bibnamefont {Weng}},\
  }\bibfield  {title} {\enquote {\bibinfo {title} {{Phase string effect in a
  doped antiferromagnet}},}\ }\href {\doibase 10.1103/PhysRevLett.77.5102}
  {\bibfield  {journal} {\bibinfo  {journal} {Physical Review Letters}\
  }\textbf {\bibinfo {volume} {77}},\ \bibinfo {pages} {5102--5105} (\bibinfo
  {year} {1996})}\BibitemShut {NoStop}%
\bibitem [{\citenamefont {Weng}\ \emph {et~al.}(1997)\citenamefont {Weng},
  \citenamefont {Sheng}, \citenamefont {Chen},\ and\ \citenamefont
  {Ting}}]{Weng1997}%
  \BibitemOpen
  \bibfield  {author} {\bibinfo {author} {\bibfnamefont {Z.~Y.}\ \bibnamefont
  {Weng}}, \bibinfo {author} {\bibfnamefont {D.~N.}\ \bibnamefont {Sheng}},
  \bibinfo {author} {\bibfnamefont {Y.-C.}\ \bibnamefont {Chen}}, \ and\
  \bibinfo {author} {\bibfnamefont {C.~S.}\ \bibnamefont {Ting}},\ }\bibfield
  {title} {\enquote {\bibinfo {title} {{Phase string effect in the $t$-$J$
  model: General theory}},}\ }\href {\doibase 10.1103/PhysRevB.55.3894}
  {\bibfield  {journal} {\bibinfo  {journal} {Physical Review B}\ }\textbf
  {\bibinfo {volume} {55}},\ \bibinfo {pages} {3894--3906} (\bibinfo {year}
  {1997})}\BibitemShut {NoStop}%
\bibitem [{\citenamefont {Weng}(2011)}]{Weng2011}%
  \BibitemOpen
  \bibfield  {author} {\bibinfo {author} {\bibfnamefont {Zheng~Yu}\
  \bibnamefont {Weng}},\ }\bibfield  {title} {\enquote {\bibinfo {title}
  {{Superconducting ground state of a doped Mott insulator}},}\ }\href
  {\doibase 10.1088/1367-2630/13/10/103039} {\bibfield  {journal} {\bibinfo
  {journal} {New Journal of Physics}\ }\textbf {\bibinfo {volume} {13}}
  (\bibinfo {year} {2011}),\ 10.1088/1367-2630/13/10/103039}\BibitemShut
  {NoStop}%
\end{thebibliography}%


\clearpage
\appendix
\widetext

\begin{center}
\textbf{\large Supplemental Material for \\``Pair density wave characterized by a hidden string order parameter''}
\end{center}

\renewcommand{\thefigure}{S\arabic{figure}}
\setcounter{figure}{0}
\renewcommand{\theequation}{S\arabic{equation}}
\setcounter{equation}{0}

This supplemental material contains more numerical results and detailed analyses to further support the conclusions we have discussed in the main text. We first introduce the general definition of the twisted quasiparticle. Then we show more data on density profiles and technical details to support the proposed phase diagram. Finally we provide more information on correlation functions and careful analyses on the composite structure of the pair-pair correlators.

\section{1. Definition of the twisted quasiparticle}
We show a simple derivation of the definition of the twisted quasiparticle $\tilde{c}_{i\sigma}$ presented in the main text. Inspired by the mutual statistics between spin and hole in the $t$-$J$ model due to the phase-string sign structure~\cite{Sheng1996,Weng1997,Wu2008}, we perform the following duality transformation~\cite{Weng2011}
\begin{equation}
    e^{i\Theta} = \text{exp}\left( -i\sum_{i\neq j} n^h_i n_{l\downarrow} \theta_{ij} \right) =\text{exp}\left({-i\sum_{i}n^h_i\Omega_{i\downarrow}}\right),
\end{equation}
where $\Omega_{i\downarrow}=\sum_{j\neq i}n_{j\downarrow}\theta_{ij}$ counts the statistical phase contributed from the $\downarrow$-spins around site $i$ and $\theta_{ij}=\operatorname{Im}\ln(z_i-z_j)$ denotes the statistical angle between site $i$ to site $j$ with $z_i=x_i+iy_i$ being the complex coordinate. Then the transformed (or twisted) fermion operator is
\begin{equation}
\begin{aligned}
    \tilde{c}_{i\sigma}&= e^{i\Theta} c_{i\sigma}e^{-i\Theta}
    =e^{-i\sum_{j\neq i}(n^h_in_{j\downarrow}\theta_{ij} +n^h_j n_{i\downarrow}\theta_{ji})} c_{i\sigma} e^{i\sum_{j\neq i}(n^h_in_{j\downarrow}\theta_{ij} +n^h_j n_{i\downarrow}\theta_{ji})}.
\end{aligned}
\end{equation}
For different spin orientations, we have
\begin{equation}\label{Eq:ctilde_updnh}
\begin{aligned}
    & \tilde{c}_{i\uparrow}
    =e^{-i\sum_{j\neq i}n^h_in_{j\downarrow}\theta_{ij}} c_{i\uparrow}
    =c_{i\uparrow}e^{-i\sum_{j\neq i}n_{j\downarrow}\theta_{ij}}
    =c_{i\uparrow}e^{-i\Omega_{i\downarrow}}, \\
    & \tilde{c}_{i\downarrow}=e^{-i\sum_{j\neq i}n^h_in_{j\downarrow}\theta_{ij} } c_{i\downarrow} e^{i\sum_{j\neq i}n^h_jn_{i\downarrow}\theta_{ji}}
    =c_{i\downarrow} e^{-i\sum_{j\neq i}(n_{j\downarrow}\theta_{ij} -n^h_j\theta_{ji})}=c_{i\downarrow} e^{-i(\Omega_{i\downarrow} + \Phi^h_i)}(-1)^{\hat{N}_h}.
\end{aligned}
\end{equation}
where $\Phi_i^h = \sum_{j\neq i} n^h_j \theta_{ij}$. The constant term $(-1)^{\hat{N}_h}$ can be omitted. According to the no-double-occupancy constraint $n_{i\uparrow} + n_{i\downarrow} + n_i^{h} = 1$, we have $\Phi_{i}^{h} = \Phi_{i}^{0} - \Omega_{i\uparrow} - \Omega_{i\downarrow}$, where  $\Phi_i^0 = \sum_{j\neq i} \theta_{ij}$ and $\Omega_{i\uparrow}=\sum_{j\neq i} n_{j\uparrow} \theta_{ij}$. Substitute this relation into Eq.~(\ref{Eq:ctilde_updnh}) and we obtain
\begin{equation}
\begin{aligned}
    & \tilde{c}_{i\uparrow} = c_{i\uparrow} e^{-i \Omega_{i\downarrow}},\quad
    \tilde{c}_{i\downarrow} = c_{i\downarrow} e^{-i \left(\Phi^0_i - \Omega_{i\uparrow} \right) }.
\end{aligned}
\end{equation}
The $\Phi_i^0$ term which contributes a trivial $2\pi$ winding flux can be omitted. Finally, we arrive at $\tilde{c}_{i\sigma}=c_{i\sigma}e^{-i\sigma\Omega_{i,-\sigma}}$. For the case of 1D hopping where $\theta_{ij}\in\{0,\pi\}$, this reduces to the transformation implemented in the main text.
Note that in the one-hole case the above string operator can be further reduced to the form used in Refs. \cite{Zhu2018,Chen2018a}.

\section{2. Technical details on processing data with U(1) symmetry}
The $t$-$J$ model under uniform Zeeman field has both charge and spin $\mathrm{U}(1)$ symmetry so that the total spin $S^z$ can only take integer values, which causes quantized $m$-$h$ curves for finite-size systems shown in Fig.~\textcolor{blue}{2} of the main text. To take the derivative $\chi=\frac{\partial m}{\partial h}$ meaningfully, we smooth such quantized curves by connecting the midpoints of those quantized plateaus and performing finite-size scaling to obtain the onset and saturation of the magnetization. In addition, the legends in Fig.~\textcolor{blue}{3}(a) and Fig.~\textcolor{blue}{4}(d) results from keeping two decimals for $m=S^z/N$ with $S^z=0,4,8,...,36$ and $S^z=0,4,8,...,56$, respectively. The continuous Fourier spectrums in Figs.~\textcolor{blue}{2}(c), (d), Figs.~\textcolor{blue}{3}(b), (d) and (f) are obtained from the discrete spectrums at $S^z=0,4,8,...,56$ by the bicubic interpolation function.

\section{3. Charge and spin density distributions}
\begin{figure}
    \centering
    \includegraphics[width=0.85\linewidth]{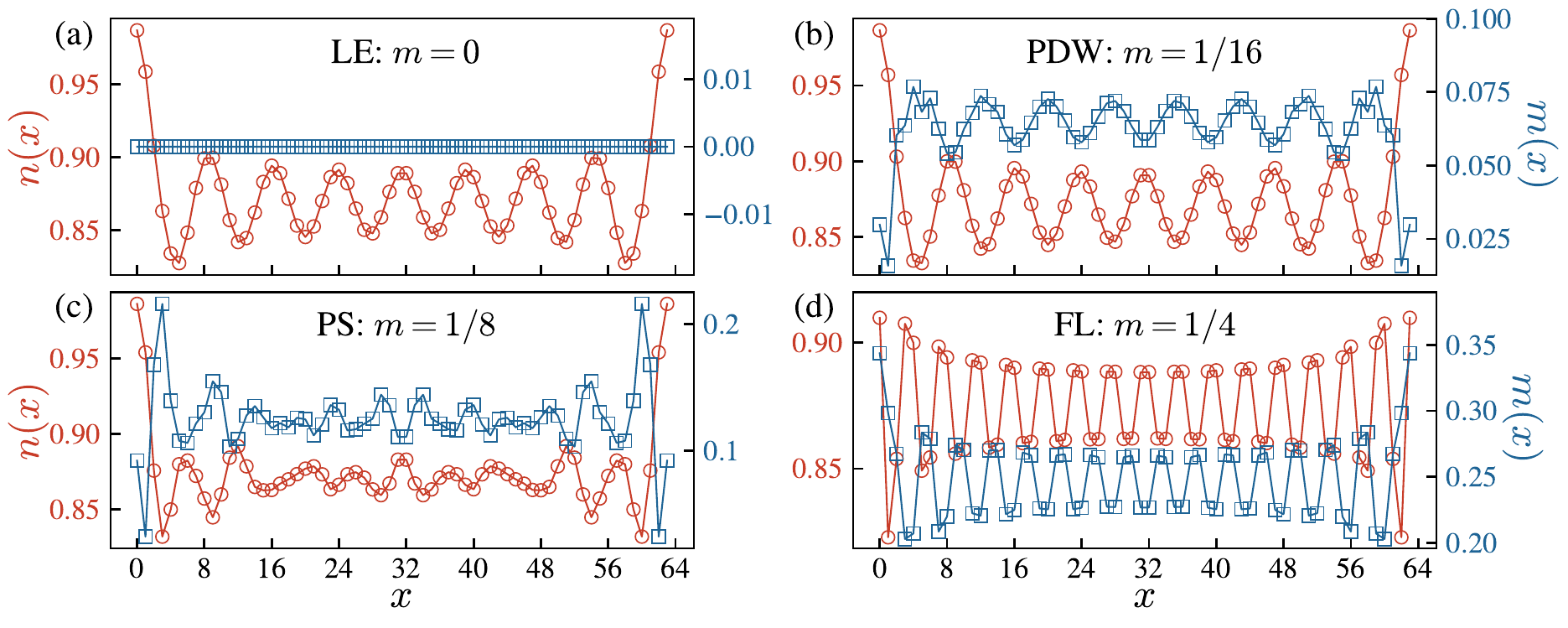}
    \caption{Typical charge density profiles $n(x)$ and spin density profiles $m(x)$ of a two-leg $t$-$J$ ladder with $\delta=1/8$, $t_\perp=3$ and lattice length $L_x=64$ with respect to the four regions of the phase diagram in the main text.}
    \label{fig:real_space_density}
\end{figure}

We provide typical real space distributions of charge and spin densities in Fig.~\ref{fig:real_space_density} as complements of the Fourier spectrums shown in Figs.~\textcolor{blue}{2}(c) and (d) of the main text, where the large uniform component $n(q=0)$ is artificially removed for better presentation. The charge and spin density distributions along the $\hat{x}$-direction are defined by $n(x)=\sum_{y}\avg{n_{x,y}}/L_y$ and $m(x)=\sum_{y}\avg{S^z_{x,y}}/L_y$, respectively. One can find that up to the boundary effect, the density profiles in the LE, PDW and FL phases have well-defined periodicities while those in the phase separation region are relatively irregular and inhomogeneous.

\begin{figure}
    \centering
    \includegraphics[width=0.72\linewidth]{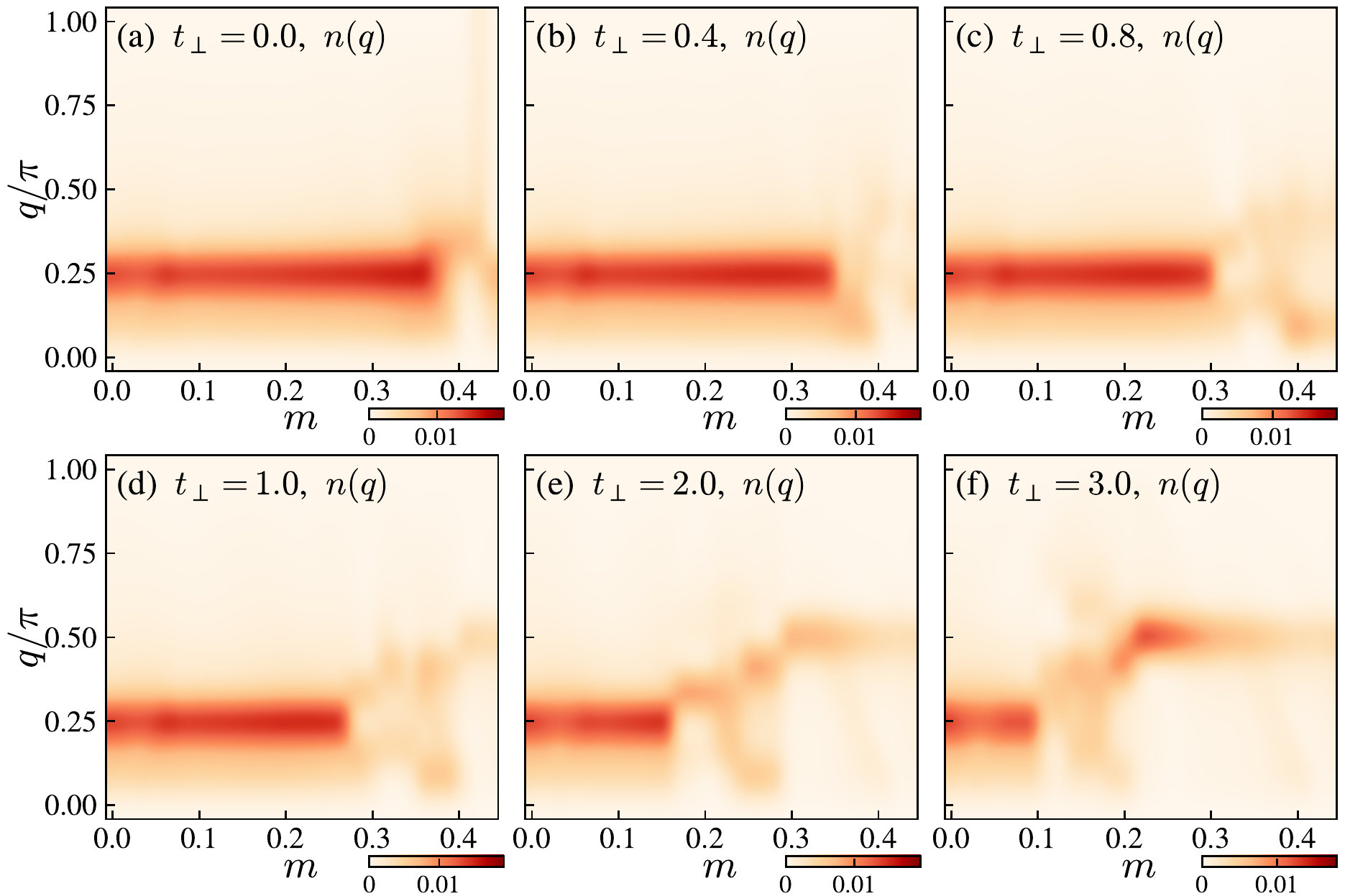}
    \caption{Fourier spectrum $n(q)$ of charge density profile vs. the total magnetization $m$ at different value of the interchain hopping integral $t_\perp$ of a two-leg $t$-$J$ ladder with doping $\delta=1/8$ and lattice length $L_x=32$.}
    \label{fig:nq_tperp}
\end{figure}

The main text contains the Fourier spectrum of charge density profile as a function of the magnetization at $t_\perp=0$ and $3$ of lattice length $L_x=64$. We depict the same quantity at more values of $t_\perp$ of length $L_x=32$ in Fig.~\ref{fig:nq_tperp}, where the discrete raw data at $S^z=0,1,2,...,28$ is interpolated. The Fourier spectrum is defined as $n(q)=\left|\sum_{x}n(x)e^{i q x}\right|/R$ where $x$ is summed over the central bulk of length $R=3L_x/4$ to reduce the boundary effect. The PDW, FL and phase separation regions together with their boundaries can be easily identified from the patterns of $n(q)$, i.e., a peak at $Q_{\rm CDW}=2\pi\delta$, a peak at $Q_{\rm CDW}=4\pi\delta$ and a relatively messy pattern with no predominant peak.

\section{4. Other correlation functions}
\begin{figure}
    \centering
    \includegraphics[width=0.85\linewidth]{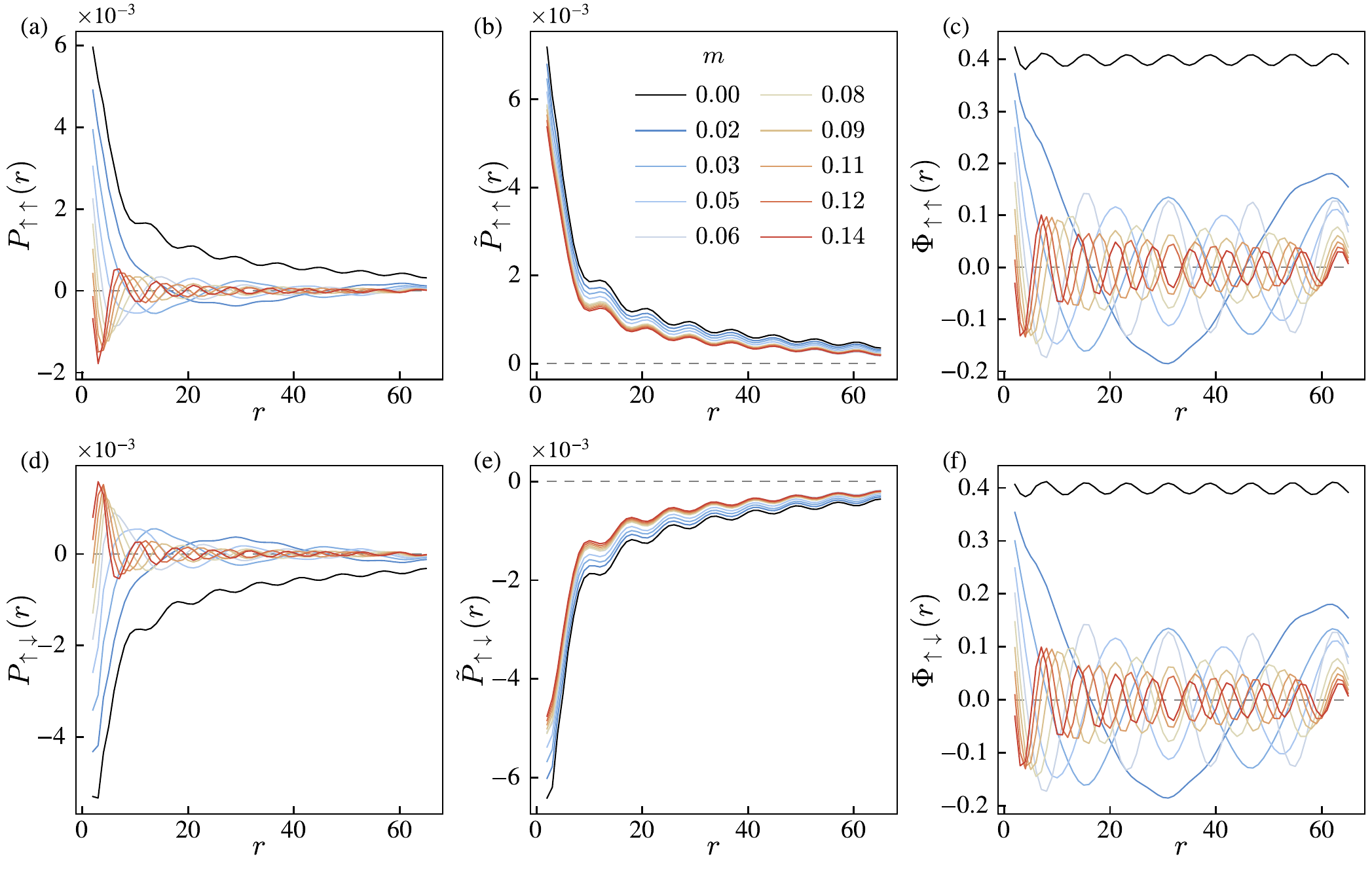}
    \caption{Four elementary correlators $P_{\sigma\sigma'}(r)$ constituting the pair-pair correlator of spin-singlets $P(r)$, together with the corresponding correlators of the twisted pairs $\tilde{P}_{\sigma\sigma'}(r)$ and the twisting phase $\Phi_{\sigma\sigma'}(r)$. These correlators are measured on a two-leg $t$-$J$ ladder with $t_\perp=0$, $\delta=1/8$ and $L_x=128$ at different magnetizations $m$. Only $P_{\uparrow\uparrow}(r)$ and $P_{\uparrow\downarrow}(r)$ are depicted since $P_{\sigma\sigma'}(r)$ and $P_{-\sigma,-\sigma'}(r)$ are equivalent under the reflection symmetry between the two chains of the two-leg ladder. }
    \label{fig:sc_r_sub_term}
\end{figure}

Fig.~\textcolor{blue}{3} depicts the pair-pair correlators of spin singlets $P(r)$, which can be represented as a summation of four elementary correlators
\begin{equation}\label{Eq:sc_sub_term}
\begin{aligned}
    \avg{\Delta_{x_0}^\dagger \Delta_{x_0+r}} &= \left\langle \frac{1}{2} \left( c_{x_02,\downarrow}^\dagger c_{x_01,\uparrow}^\dagger -  c_{x_02,\uparrow}^\dagger c_{x_01,\downarrow}^\dagger \right)\left( c_{(x_0+r)1,\uparrow}c_{(x_0+r)2,\downarrow} - c_{(x_0+r)1,\downarrow}c_{(x_0+r)2,\uparrow} \right) \right\rangle \\
    &= - \frac{1}{2} \Big( \avg{ c_{x_01,\uparrow}^\dagger c_{x_02,\downarrow}^\dagger c_{(x_0+r)1,\uparrow} c_{(x_0+r)2,\downarrow} } - \avg{ c_{x_01,\uparrow}^\dagger c_{x_02,\downarrow}^\dagger c_{(x_0+r)1,\downarrow} c_{(x_0+r)2,\uparrow} } \\
    &\quad\quad~ - \avg{ c_{x_01,\downarrow}^\dagger c_{x_02,\uparrow}^\dagger c_{(x_0+r)1,\uparrow} c_{(x_0+r)2,\downarrow} } + \avg{ c_{x_01,\downarrow}^\dagger c_{x_02,\uparrow}^\dagger c_{(x_0+r)1,\downarrow} c_{(x_0+r)2,\uparrow} } \Big).
\end{aligned}
\end{equation}
The overall negative sign before the $1/2$ factor is omitted in Fig.~\textcolor{blue}{3} by convention. We denote these elementary correlators as $P_{\sigma\sigma'}(r)=\avg{c_{x_01,\sigma}^\dagger c_{x_02,-\sigma}^\dagger c_{(x_0+r)1,\sigma'} c_{(x_0+r)2,-\sigma'}}$. The corresponding correlators of the pairing amplitude and phase can be defined similarly, i.e., $\tilde{P}_{\sigma\sigma'}(r)=\avg{\tilde{c}_{x_01,\sigma}^\dagger \tilde{c}_{x_02,-\sigma}^\dagger \tilde{c}_{(x_0+r)1,\sigma'} \tilde{c}_{(x_0+r)2,-\sigma'}}$ and $\Phi_{\sigma\sigma'}(r)=\avg{e^{-i\phi_{x_0,\sigma}} e^{i\phi_{x_0+r,\sigma'}}}$, where $\phi_{x,\sigma}=\Omega_{x1,-\sigma}+\Omega_{x2,\sigma}$. ($\phi_{x,\uparrow}$ is exactly the string-like operator $\phi_{x}$ in the main text.) To illustrate the composite structure of the pairing amplitude and phase more directly, we depict these elementary correlators in Fig.~\ref{fig:sc_r_sub_term} on top of the summed correlators $P(r)$ in the main text. In consistent with the conclusions in the main text, one can find that $\tilde{P}_{\sigma\sigma'}(r)$ still exhibits no sign change and $\Phi_{\sigma\sigma'}(r)$ captures the PDW oscillation precisely. 

Moreover, $\Phi_{\sigma\sigma}(r)$ and $\Phi_{\sigma,-\sigma}(r)$ [$P_{\sigma\sigma}(r)$ and $P_{\sigma,-\sigma}(r)$, $\tilde{P}_{\sigma\sigma}(r)$ and $\tilde{P}_{\sigma,-\sigma}(r)$] almost equal to each other [up to an overall negative sign], 
indicating that $\Phi(r)$ as a common factor can be extracted out from the summation in Eq.~(\ref{Eq:sc_sub_term}). This explains why the summed correlator $P(r)$ also has a well-defined composite structure $P(r)\propto \tilde{P}(r)\Phi(r)$.

In addition, it is worth mentioning that in the LE and PDW phases, the overall magnitude of pair-pair correlator of spin singlets is notably higher than those of other spatial or spin configurations, e.g., bonds along the $\hat{x}$-direction, and spin triplets of both $S=0$ and $S=1$. 

Finally, we provide the estimation of typical spatial decay behaviors of the pair-pair correlator together with other correlation functions in Table.~\ref{table:scaling_exponent}, including the density-density correlation function $D(r)=\frac{1}{L_y}\sum_{y=1}^{L_y} \big(\avg{n_{x_0y} n_{(x_0+r)y}}-\avg{n_{x_0y}}\avg{n_{(x_0+r)y}}\big)$, the single-particle Green's function $G_\sigma(r)=\frac{1}{L_y}\sum_{y=1}^{L_y} \avg{c_{x_0y,\sigma}^\dagger c_{(x_0+r)y,\sigma}}$ and the spin-spin correlation function $F_{\alpha}(r)=\frac{1}{L_y}\sum_{y=1}^{L_y} \big(\avg{S_{x_0y,\sigma}^\alpha S_{(x_0+r)y,\sigma}^\alpha}-\avg{S_{x_0y,\sigma}^\alpha }\avg{S_{(x_0+r)y,\sigma}^\alpha}\big)$ where $\alpha\in\{x,y,z\}$ is the index of the spin component. One can see that $P(r)$ becomes short-ranged and $G_{\sigma}(r)$ becomes quasi-long-ranged from the LE and PDW regimes to the PS and FL regimes, in accordance with the results of the binding energy. $F_{x}(r)$ becomes quasi-long-ranged from the LE phase to the PDW phase, which further reveals the distinction between these two regimes.

\begin{table*}
    \centering
	\caption{Estimation of typical power-law exponents or correlation lengths of the pair-pair correlator $P(r)$, the density-density correlator $D(r)$, the single-particle Green's function $G_{\sigma}(r)$ and the spin-spin correlator $F_{\alpha}(r)$. If the correlators decay algebraically $\sim r^{-K}$, we denote the corresponding power-law exponents as $K_{sc}$, $K_{c}$, $K_{G}$ and $K_s$, respectively. Otherwise if they decay exponentially $\sim e^{-r/\xi}$, we denote the correlation lengths as $\xi_{sc}$, $\xi_{c}$, $\xi_{G}$ and $\xi_s$.}
	\label{table:scaling_exponent}
    \begin{tabular}{c|ccccc}
        \hline
        \hline
        \quad Phase \quad & \quad\quad $P(r)$ \quad\quad & \quad\quad $D(r)$ \quad\quad & \quad\quad $G_{\sigma}(r)$ \quad\quad & \quad\quad $F_{x}(r)$ \quad\quad & \quad\quad $F_{z}(r)$ \quad\quad \\ \hline
        LE & $K_{sc} \lesssim 1$ & $ 1< K_{c} < 2$ & $\xi_{G}< 5$ & $\xi_{s}< 7$ & $\xi_{s}< 7$ \\ \hline
        PDW & $1\lesssim K_{sc} \lesssim 2$ & $K_{c} \lesssim 1$ & $\xi_{G}< 7$ & $K_{s} < 1$ & $\xi_{s}< 7$ \\ \hline
        PS & $\xi_{sc}< 4$ & $K_{c} < 2$ & $K_{G} < 2$ & $K_{s} < 1$ & $K_{s}< 2$ \\ \hline
        FL & $\xi_{sc}< 4$ & $K_{c} < 2$ & $K_{G} < 2$ & $K_{s} < 1$ & $K_{s}< 2$ \\ \hline
    \end{tabular}
	
\end{table*}

\section{5. The accurate separation of the pairing amplitude and phase at $t_{\perp}=0$}
\begin{figure}
    \centering
    \includegraphics[width=0.85\linewidth]{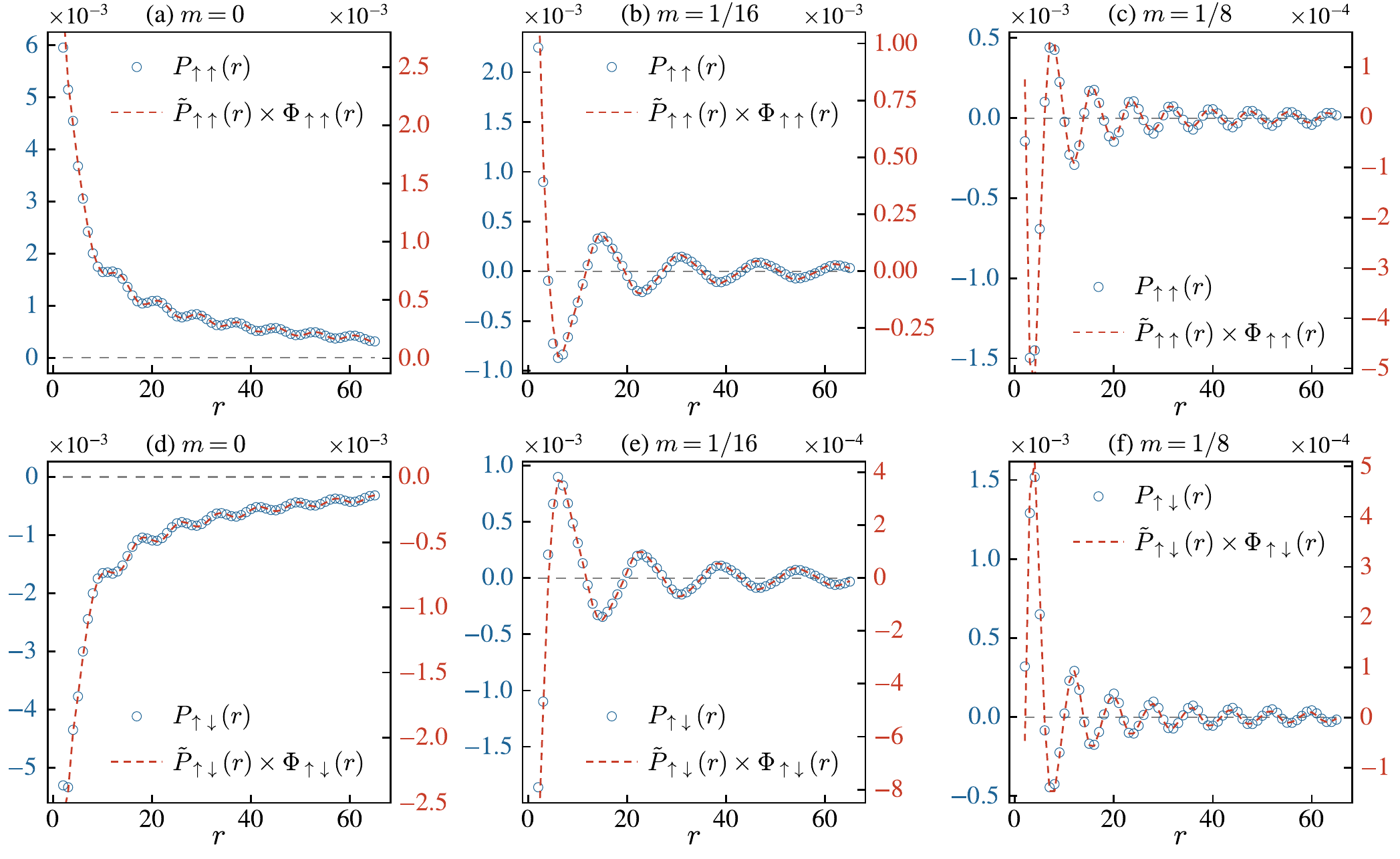}
    \caption{Recombining $\tilde{P}_{\sigma\sigma'}(r)$ and $\Phi_{\sigma\sigma'}(r)$ can reproduce $P_{\sigma\sigma'}(r)$ very accurately except for an overall factor. These correlators are measured on a two-leg $t$-$J$ ladder with $t_\perp=0$, $\delta=1/8$ and $L_x=128$.}
    \label{fig:sc_r_recombine}
\end{figure}

We may further check the composite structure of the pairing order parameter in Eq.~(\textcolor{blue}{4}) of the main text by multiplying the correlators of $\tilde{P}_{\sigma\sigma'}(r)$ and $\Phi_{\sigma\sigma'}(r)$ in comparison  with the original pair-pair correlator $P_{\sigma\sigma'}(r)$ in Fig.~\ref{fig:sc_r_recombine}. One can find that up to an overall factor, the recombined $\tilde{P}_{\sigma\sigma'}(r)\times \Phi_{\sigma\sigma'}(r)$ coincides very precisely with $P_{\sigma\sigma'}(r)$, which confirms that the pairing amplitude and phase can be well described by $\avg{\tilde{c}_{x1,\uparrow}\tilde{c}_{x2,\downarrow}}$ and $\avg{e^{i\phi_x}}$ respectively, i.e., $\avg{c_{x1,\uparrow}c_{x2,\downarrow}} = \avg{\tilde{c}_{x1,\uparrow}\tilde{c}_{x2,\downarrow} e^{i\phi_x}} \propto \avg{\tilde{c}_{x1,\uparrow}\tilde{c}_{x2,\downarrow}} \avg{e^{i\phi_x}}$ as a generalized mean-field-type separation (with an appropriately determined overall renormalization factor). 

\end{document}